\documentclass[aps,preprint,nofootinbib]{revtex4}%
\usepackage{amsmath}
\usepackage{graphicx}
\usepackage{amsfonts}
\usepackage{amssymb}%
\providecommand{\U}[1]{\protect\rule{.1in}{.1in}}
\providecommand{\U}[1]{\protect\rule{.1in}{.1in}}
\providecommand{\U}[1]{\protect\rule{.1in}{.1in}}

\begin{document}
\preprint{ }
\title{}
\author{}
\maketitle

\begin{center}
\rightline{USC-HEP-10B2}

{\Large The Big Bang and Inflation United}

{\Large by an Analytic Solution }\footnote{Work partially supported by the US
Department of Energy, grant number DE-FG03-84ER40168.}

{\vskip0.30cm}

\textbf{Itzhak Bars}$^{\ast}$\textbf{ and Shih-Hung Chen}$^{\dagger}$\textbf{
}

{\vskip0.3cm}

$^{\ast}$\textsl{Department of Physics and Astronomy}

\textsl{University of Southern California,\ Los Angeles, CA 90089-2535 USA}

\medskip

$^{\dagger}$\textsl{Department of Physics and School of Earth and Space
Exploration}

\textsl{Arizona State University, Tempe, AZ 85287-1404 USA}

{\vskip0.5cm} \textbf{Abstract}
\end{center}

Exact analytic solutions for a class of scalar-tensor gravity theories with a
hyperbolic scalar potential are presented. Using an exact solution we have
successfully constructed a model of inflation that produces the spectral
index, the running of the spectral index and the amplitude of scalar
perturbations within the constraints given by the WMAP 7 years data. The model
simultaneously describes the Big Bang and inflation connected by a specific
time delay between them so that these two events are regarded as dependent on
each other. In solving the Fridemann equations, we have utilized an essential
Weyl symmetry of our theory in 3+1 dimensions which is a predicted remaining
symmetry of 2T-physics field theory in 4+2 dimensions. This led to a new
method of obtaining analytic solutions in 1T field theory which could in
principle be used to solve more complicated theories with more scalar fields.
Some additional distinguishing properties of the solution includes the fact
that there are early periods of time when the slow roll approximation is not
valid. Furthermore, the inflaton does not decrease monotonically with time,
rather it oscillates around the potential minimum while settling down, unlike
the slow roll approximation. While the model we used for illustration purposes
is realistic in most respects, it lacks a mechanism for stopping inflation.
The technique of obtaining analytic solutions opens a new window for studying
inflation, and other applications, more precisely than using
approximations.\bigskip



\newpage

\section{Introduction}

\label{intro}

Most inflation theories involve one or more scalar fields which are called
inflatons \cite{Guth}\cite{chaotic inflation}\cite{Steinhardt}%
\cite{Steinhardt2}. The slow-roll approximation is a standard technique used
in the study of inflation generated by different inflaton potentials. However,
for the slow-roll approximation to be valid, the shape of the inflaton
potential has to be shallow. This is because only with a shallow inflaton
potential the kinetic energy of the inflaton can be neglected compared to the
potential energy. \ Therefore the shape of inflaton potentials have been
restricted in the past to apply the slow-roll approximation. This
approximation cannot be used if we want to figure out the dynamics of inflaton
fields in regions where the kinetic energy cannot be neglected. To obtain
solutions in these regions, one has to solve the full second order coupled
nonlinear differential equations. In general, these kinds of equations are
difficult to solve and are approached by numerical methods. \ 

In this paper, we will analytically solve a scalar-tensor theory with a scalar
field $\sigma\left(  x\right)  $ minimally coupled to gravity. The full action
of our theory is a standard scalar-tensor theory of gravity with a single
scalar%
\begin{equation}
S=\int d^{4}x\sqrt{-g}\left\{  \frac{1}{2\kappa^{2}}R-\frac{1}{2}g^{\mu\nu
}\partial_{\mu}\sigma\partial_{\nu}\sigma-V\left(  \sigma\right)  \right\}
\label{theory}%
\end{equation}
For the present application, $\sigma$ will be the inflaton, but in other
applications of our method it may have other interpretations. The potential
is
\begin{equation}
V\left(  \sigma\right)  =\left(  \frac{6}{\kappa^{2}}\right)  ^{2}\left(
c\sinh^{4}\left(  \sqrt{\frac{\kappa^{2}}{6}}\sigma\right)  +b\cosh^{4}\left(
\sqrt{\frac{\kappa^{2}}{6}}\sigma\right)  \right)  \label{inflaton potential}%
\end{equation}
where $b$ and $c$ are dimensionless free parameters of the potential, and
$\frac{1}{\kappa}$ is the reduced Planck mass $\widetilde{m}_{p}=\frac
{1}{\kappa},$ where $\widetilde{m}_{p}=\frac{m_{p}}{\sqrt{8\pi}}%
=2.43\times10^{18}GeV$. For our illustrative model we will further specialize
to $c=64b,$ and fix some integration constant to a specific value ($\delta=0,$
see below), for no particular reason other than plotting some graphs. With
this choice of parameters we will obtain a model of inflation that matches all
the current observational constraints \cite{Kinney WMAP 5years}\cite{WMAP
7years}\cite{amplitude of P_R}.

Our main purpose here is to illustrate the method of analytic computation, as
well as the underlying important ideas inspired by 2T-physics that lead to
this method and potential. The essential ingredients that permit us to obtain
an exact solution is a reformulation of the above theory as a gauge fixed
version of a theory with two conformally coupled scalars $\phi,s$. This
structure includes an essential local scaling symmetry, or Weyl symmetry, that
reduces the theory to one degree of freedom while also generating the Newton
constant from an initial scale invariant theory with no dimensionful
constants. This symmetry structure in 3+1 dimensions is a direct outcome of
2T-gravity \cite{2T gravity} from which the Weyl symmetry emerges as a
left-over of general coordinate reparametrizaton symmetry in 4+2 dimensions.
The surviving Weyl symmetry can be gauge fixed in different ways. One gauge
choice gives the theory defined above which appears very difficult to solve.
Another gauge choice, leads to fully decoupled equations for two scalars, for
the special potential given above, for the case of any conformally flat
metric, such as the Robertson-Walker case. These structures and the method of
solution will be explained in more detail in section (\ref{solving}).

The observational constraints on inflation include the value of the scalar
spectral index, the tensor to scalar ratio and the amplitude of scalar
perturbation. Following the definition of the potential-slow-roll
approximation given by Liddle and Lyth \cite{slow roll definition1}\cite{slow
roll definition2}, the validity of the slow-roll approximation for some
potential $V\left(  \sigma\right)  $ requires two slow-roll parameters
$\varepsilon_{V},\eta_{V}$ to be much smaller than $1$
\begin{equation}
\varepsilon_{V}\equiv\frac{1}{2\kappa^{2}}\left(  \frac{V^{\prime}\left(
\sigma\right)  }{V\left(  \sigma\right)  }\right)  ^{2}\ll1\text{ and
}\left\vert \eta_{V}\right\vert \equiv\left\vert \frac{1}{\kappa^{2}}%
\frac{V^{\prime\prime}\left(  \sigma\right)  }{V\left(  \sigma\right)
}\right\vert \ll1. \label{slow}%
\end{equation}
It turns out that our model in Eqs.(\ref{inflaton potential},\ref{theory})
requires the inflaton to evolve in a period of time from the Big Bang to well
into the period of inflation when the slow-roll approximation does not apply.
During this time, there are several interesting behaviors of the exact
solutions that cannot be captured by the slow-roll approximation, including
the connection between the Big Bang and inflation, and an oscillatory behavior
of the inflaton field as it settles down, which could not be discussed before.
After some time well into inflation, the kinetic energy decreases
monotonically, thus the equation of state asymptotically approaches
$w\rightarrow-1$ at late times.

Exact solutions for several different inflation potentials have been reported
\cite{exact solutions1}\cite{exact solutions2}\cite{Easther}\cite{Easther2}%
\cite{Barrow}\cite{Barrow2}, but none of them fall into the category we
discuss in this paper. There are several interesting properties of our model.

\begin{itemize}
\item First, if we trace back the dynamics of the cosmological scale factor
$a\left(  \tau\right)  $ analytically, there is a specific time when it
vanishes $a\left(  \tau_{BB}\right)  =0$ thereby defining the Big Bang. This
point in conformal time $\tau=\tau_{BB}$ is the definition of beginning of
physical \ \textquotedblleft time\textquotedblright\ $t\left(  \tau
_{BB}\right)  =0$ as defined by a comoving observer.

\item Second, our exact solution simultaneously captures analytically both the
Big Bang and inflation. In particular, inflation does not happen right after
the Big Bang, rather it is determined by the model that it happens around
$10^{5}$ Planck times $t_{I}\sim10^{5}\,t_{Planck}$ after the Big Bang.
Furthermore, unlike usual practice in inflation models, we do not artificially
insert a boundary value for the inflaton $\sigma\left(  \tau\right)  $, rather
it is predicted to start from infinity at the Big Bang and then smoothly
connect to the inflation period.

\item Third, unlike most inflation theories where the inflaton field
$\sigma\left(  \tau\right)  $ decreases monotonically during inflation, after
it drops from its infinite value at the Big Bang, it oscillates around the
potential minimum during inflation. So, by contrast to widely used
approximations in past analyses \cite{Kinney WMAP 5years}, in our model time
cannot be exchanged with field strength, showing again that there is no
substitute for our type of exact analytic solutions.
\end{itemize}

Among these properties, the first one is tentative, since it is not clear we
can apply a purely classical theory to the very early moment of the universe
where quantum effects are expected to play an important role. But in any case
we think that the interesting physical picture of the relationship between the
Big Bang and inflation may still remain as a physically correct feature. On
the opposite end, inflation in our model does not end by itself, rather we
need to introduce other mechanisms to end inflation around $10^{7}$ Planck
times after the Big Bang. The mechanism of stopping inflation requires
additional physical features involving reheating that may be possible to
incorporate naturally in a more complete physical theory. For this reason we
leave it to future investigations, while in this paper the primary focus is on
introducing a technique for obtaining exact solutions for the scalar-tensor
theory with the potential $\left(  \ref{inflaton potential}\right)  $ and
showing that it provides an attractive description of the Big-Bang and
inflation$.$

The paper is organized as follows: in section (\ref{constraints}) we review
the phenomenological constraints of inflation theories. In section
(\ref{solving}) we explain in detail a technique for obtaining analytic
solutions which we adopt from our study of 2T-gravity. This enabled us to
obtain analytic solutions for the standard gravity action in $\left(
\ref{theory}\right)  .$ In section (\ref{toy model}) we construct a model of
inflation that matches all the observational constraints. In section
(\ref{conclusion}) we summarize our conclusions and point out some future directions.

\section{Phenomenological constraints of inflation models}

\label{constraints}

The latest observational constraints on inflation models from WMAP 7 years
data can be found in\ ref \cite{WMAP 7years}. In this section we will briefly
review the zeroth order and first order properties of inflation theories.

First, an inflation theory must be able to generate a period of inflation
which is defined as $\frac{d^{2}a_{E}}{dt^{2}}>0$ where $a_{E}\left(
t\right)  $ is the scale factor of the flat Friedmann-Roberson-Walker (FRW)
metric in the Einstein frame%
\begin{align}
ds^{2}  &  =-dt^{2}+a_{E}^{2}\left(  t\right)  \left(  dx^{2}+dy^{2}%
+dz^{2}\right) \nonumber\\
&  =a_{E}^{2}\left(  \tau\right)  \left(  -d\tau^{2}+dx^{2}+dy^{2}%
+dz^{2}\right)  \label{FRW}%
\end{align}
In the second line we have pulled out a common factor $a_{E}^{2}\left(
\tau\right)  \equiv a_{E}^{2}\left(  t\left(  \tau\right)  \right)  $ and
defined the conformal time $\tau,$ with $t=t\left(  \tau\right)  $ given by
the relation $a_{E}\left(  \tau\right)  d\tau=dt.$ Using conformal time, the
derivative with respect to time $t$ can be rewritten as $\frac{d}{dt}=\frac
{1}{a_{E}}\frac{d}{d\tau},$ therefore, the Hubble parameter $H\equiv\left[
\left(  \frac{da_{E}}{dt}\right)  /a_{E}\right]  $ can be expressed as
\begin{equation}
H\left(  \tau\right)  =\frac{\dot{a}_{E}\left(  \tau\right)  }{a_{E}%
^{2}\left(  \tau\right)  },
\end{equation}
where the overdot denotes derivative with respect to conformal time. Second,
an inflation theory should produce more than 60 e-folds of expansion in order
to solve the horizon and the flatness problems, which means $\ln\left(
\frac{a_{E}^{end}}{a_{E}^{begin}}\right)  >60,$ where $a_{E}^{end}$ and
$a_{E}^{begin}$ are the scale factors at the end and the beginning of
inflation respectively. The above two constraints are the zeroth order
properties of an inflation theory that comes from a purely classical gravity theory.

To include the fluctuations of the fields that cause the primordial
perturbation, one should match the amplitude of fluctuations at the horizon
crossing to the current observational anisotropy of the CMB. This is explained
in detail in \cite{Baumann TASI09},\cite{Lyth},\cite{Lyth Riotto}%
,\cite{Dodelson book}, while here we will outline the procedure. We consider
the small oscillations for the perturbed FRW metric and the inflaton%

\begin{align}
ds^{2}  &  =a_{E}^{2}\left(  \tau\right)  \left\{  -(1+2A)d\tau^{2}%
-2\partial_{i}Bdx^{i}d\tau+\left[  \left(  1+2\mathcal{R}\right)  \delta
_{ij}+\partial_{i}\partial_{j}H_{T}\right]  dx^{i}dx^{j}\right\} \\
\sigma &  =\sigma_{E}\left(  \tau\right)  +\delta\sigma
\end{align}
In this paper we will provide exact analytic solutions for the backgrounds
$a_{E}\left(  \tau\right)  $ and $\sigma_{E}\left(  \tau\right)  .$ For the
perturbations we choose a coordinate reparametrization gauge in which
$\delta\sigma=0,$ and concentrate on the relevant curvature perturbation $R$.
To compute the amplitude of this perturbation, one first expands the action in
Eq.$\left(  \ref{theory}\right)  $ up to second order and then puts the
background evolution on shell. The resulting action for the perturbation is%
\begin{equation}
S_{\left(  2\right)  }=\frac{1}{2}\int d\tau d^{3}x~\frac{\dot{\sigma}_{E}%
^{2}}{H^{2}}\left[  \mathcal{\dot{R}}^{2}-\left(  \partial_{i}\mathcal{R}%
\right)  ^{2}\right]
\end{equation}
After defining a rescaled amplitude $v\equiv zR,$ using the factor
\begin{equation}
z\left(  \tau\right)  \equiv\frac{\dot{\sigma}_{E}\left(  \tau\right)
}{H\left(  \tau\right)  }%
\end{equation}
that depends on background evolution, the action for the perturbation
$v\left(  \tau,\vec{x}\right)  $ becomes%
\begin{equation}
S_{\left(  2\right)  }=\frac{1}{2}\int d\tau d^{3}x\left[  \left(  \dot
{v}\right)  ^{2}-\left(  \partial_{i}v\right)  ^{2}+\frac{\overset{\cdot\cdot
}{z}}{z}v^{2}\right]  .
\end{equation}
The Fourier transform of $v\left(  \tau,\vec{x}\right)  $ defines the mode
function $v_{k}\left(  \tau\right)  $ in momentum space%
\begin{equation}
v\left(  \tau,\vec{x}\right)  =\int\frac{d^{3}x}{\left(  2\pi\right)  ^{3}%
}\left(  a_{k}v_{k}\left(  \tau\right)  e^{i\vec{k}\cdot\vec{x}}%
+a_{k}^{\dagger}v_{k}^{\ast}\left(  \tau\right)  e^{-i\vec{k}\cdot\vec{x}%
}\right)  .
\end{equation}
It satisfies the Mukhanov-Sasaki equation which is obtained by minimizing
$S_{\left(  2\right)  }$
\begin{equation}
\overset{\cdot\cdot}{v_{k}}+\left[  k^{2}-\frac{\overset{\cdot\cdot}{z}}%
{z}\right]  v_{k}=0. \label{muk}%
\end{equation}
This equation is to be solved along with a physical boundary condition that
corresponds to choosing a particular vacuum. Then one can compute the power
spectrum $P_{\mathcal{R}}\left(  k\right)  $, the spectral index $n_{s}\left(
k\right)  $ and the running of the spectral index $n^{\prime}\left(  k\right)
$ as follows
\begin{align}
&  P_{\mathcal{R}}\left(  k\right)  \equiv\frac{\left\vert v_{k}(\tau_{\ast
}\left(  k\right)  )\right\vert ^{2}}{z^{2}\left(  \tau_{\ast}\left(
k\right)  \right)  };\text{ }\tau_{\ast}\left(  k\right)  \text{ given by
}a_{E}\left(  \tau_{\ast}\right)  H\left(  \tau_{\ast}\right)
=k,\label{scalar power}\\
&  n_{s}\left(  k\right)  -1\equiv\frac{d\ln(k^{3}P_{\mathcal{R}}\left(
k\right)  )}{d\ln k},\;\;n^{\prime}\left(  k\right)  \equiv\frac{dn_{s}\left(
k\right)  }{d\ln k}. \label{spo2}%
\end{align}
Notice that all the above quantities are evaluated at horizon crossing
($k=a_{E}H$) which defines the time $\tau_{\ast}\left(  k\right)  .$ This is
because the curvature perturbation is \ freezed out when the wave length $1/k$
stretches outside the horizon. In general, $\frac{\overset{\cdot\cdot}{z}}{z}$
(given below) is a complicated function of $\tau\,$that renders the analytic
solution to Eq.$\left(  \ref{muk}\right)  $ difficult or impossible.

The solution of Eq.(\ref{muk}) has commonly been discussed for the cases in
which the background $a_{E}\left(  \tau\right)  ,\sigma_{E}\left(
\tau\right)  $ can be approximated by the de Sitter background and the slow
roll approximations. The factor $\frac{\overset{\cdot\cdot}{z}}{z}$ can then
be approximated by $\frac{\alpha}{\tau^{2}}$ for which Eq.(\ref{muk}) is
solvable analytically. However, in our model, the behavior of $\frac
{\overset{\cdot\cdot}{z}}{z}$ is far from this common fit function
$\frac{\alpha}{\tau^{2}}$. Hence we suggest a different fit function to
approximate $\frac{\overset{\cdot\cdot}{z}}{z}$ in the relevant time period
close to $\tau\sim\tau_{\ast}\left(  k\right)  $ so that this equation is
solvable analytically.

The tensor perturbation $h_{ij}$ is discussed in a similar way, starting with
\begin{equation}
ds^{2}=a_{E}^{2}\left(  -d\tau^{2}+\left(  \delta_{ij}+h_{ij}\left(  \tau
,\vec{x}\right)  \right)  dx^{i}dx^{j}\right)  .
\end{equation}
The action for this perturbation is
\begin{equation}
S_{\left(  2\right)  }=\frac{1}{\kappa^{2}}\int d\tau d^{3}x~a_{E}^{2}\left(
\tau\right)  \left[  \dot{h}_{ij}^{2}-\left(  \partial_{k}h_{ij}\right)
^{2}\right]
\end{equation}
The Fourier transform of $h_{ij}\left(  \tau,\vec{x}\right)  $ is
$a^{s}\left(  k\right)  \varepsilon_{ij}^{s}\left(  k\right)  h_{k}^{s}\left(
\tau\right)  e^{i\vec{k}\cdot\vec{x}}$ plus the hermitian conjugate, where
$\varepsilon_{ij}^{s}\left(  k\right)  $ is the spin-two polarization tensor
and $s$ denotes the polarization. The rescaled amplitude defined by $\mu
_{k}^{s}\equiv\frac{1}{2}a_{E}h_{k}^{s}$ is then governed by the following
action for the mode $\mu_{k}^{s}\left(  \tau\right)  $
\begin{equation}
S_{\left(  2\right)  }=\frac{1}{2\kappa^{2}}\sum_{s}\int d\tau d^{3}k\left[
\left(  \dot{\mu}_{k}^{s}\right)  ^{2}-\left(  k^{2}-\frac{\ddot{a}_{E}}%
{a_{E}}\right)  \left(  \mu_{k}^{s}\right)  ^{2}\right]  .
\end{equation}
In our model the behavior of $\frac{\ddot{a}_{E}}{a_{E}}$ is similar to the
standard case and can be approximated by $\alpha/\tau^{2}$, so the Mukhanov
equation for $\mu_{k}^{s}\left(  \tau\right)  $ can be solved in the usual
way. As in the scalar case, we can then compute the power spectrum for the
tensor, $P_{T}\left(  k\right)  ,$ at horizon crossing. Finally, another
observational quantity, the tensor to scalar ratio $r$ is defined as%
\begin{equation}
r\equiv\frac{P_{T}\left(  k\right)  }{P_{\mathcal{R}}\left(  k\right)
}.\text{ } \label{r}%
\end{equation}

The phenomenologically allowed ranges for $r$ and $n$ and $n^{\prime}$ are
plotted in \cite{Kinney WMAP 5years}\cite{WMAP 7years}. A pure power law
spectrum should predict zero running $n^{\prime}\sim0$. Finally, the amplitude
of scalar perturbations is also observed to be around $P_{\mathcal{R}}\left(
k\right)  \sim10^{-5}$ \cite{amplitude of P_R}.

In this paper we construct a model of inflation that satisfies all of the
above observational constraints.

\section{Solving the theory analytically}

\label{solving}

In this section we will solve analytically the equations of motion for the
action $\left(  \ref{theory}\right)  .$ Assuming the scalar field $\sigma$ is
homogeneous in space and using the flat Friedmann-Roberson-Walker metric
$\left(  \ref{FRW}\right)  $, there are two well known independent Einstein
equations and one equation for $\sigma$ \cite{Friedmann papaer}.%
\begin{gather}
\frac{\dot{a}_{E}^{2}}{a_{E}^{4}}=\frac{\kappa^{2}}{3}\left[  \frac{1}%
{2a_{E}^{2}}\overset{\cdot}{\sigma}^{2}+V\left(  \sigma\right)  \right]
\label{00}\\
\frac{\overset{\cdot\cdot}{a_{E}}}{a_{E}^{3}}-\frac{\dot{a}_{E}^{2}}{a_{E}%
^{4}}=-\frac{\kappa^{2}}{3}\left[  \frac{1}{a_{E}^{2}}\overset{\cdot}{\sigma
}^{2}-V\left(  \sigma\right)  \right] \label{11}\\
\frac{\overset{\cdot\cdot}{\sigma}}{a_{E}^{2}}+2\frac{\dot{a}_{E}}{a_{E}^{3}%
}\overset{\cdot}{\sigma}+V^{\prime}\left(  \sigma\right)  =0 \label{eom sigma}%
\end{gather}
Where prime represents the derivative with respect to $\sigma.$ The first two
equations are the $\mu\nu=00$ and $\mu\nu=11$ components of Einstein Equations
and the third equation is the equation of motion for $\sigma,$ all expressed
in terms of the conformal time $\tau.$ These three equations are coupled
second order nonlinear differential equations. We will obtain all the exact
solutions of these equations as displayed in Eq.(\ref{AE}) and
Eq.(\ref{sigma as function of z and t}) when the potential is as given in
Eq.(\ref{inflaton potential}). To obtain this solution we will use a technique
which we developed in the context of 2T-gravity that results in a theory with
Weyl symmetry when reduced to 1T shadow. For this reason we give a brief
outline of how this is inspired from 2T-gravity.

\subsection{The 2T approach to ordinary gravity}

Here we will not discuss 2T-gravity itself \cite{2T gravity}\cite{geometry 2T
gravity}, which is a theory in $d+2$ dimensions. We will only mention the
crucial property of this formulation, namely that it has the right mix of
gauge symmetries to eliminate all ghosts from the 2T fields (including those
extra timelike components in vector or tensor fields) and yield shadow fields
in two lower dimensions that are ghost free fields in physical interacting 1T
field theories in $d$ dimensions. Dualities relate the many possible shadow 1T
field theories that emerge in the process of gauge fixing. For our purposes
here we concentrate only on the so called \textquotedblleft conformal
shadow\textquotedblright. The conformal shadow (like other shadows as well) in
$d$ dimensions captures holographically all the gauge invariant content of the
2T-gravity parent theory in $d+2$ dimensions \cite{GMtalk}.

The action for the conformal shadow of \textit{pure} 2T-gravity yields
ordinary 1T General Relativity in $d$ dimensions with some constraints imposed
on it. In particular it contains a dilaton $\phi$ and the full action has the
form
\begin{equation}
S_{grav}=\int d^{d}x\sqrt{-g}\left\{  z_{d}\phi^{2}R\left(  g\right)
+\frac{1}{2}g^{\mu\nu}\partial_{\mu}\phi\partial_{\nu}\phi-V\left(
\phi\right)  \right\}  ,
\end{equation}
where $\phi\left(  x\right)  ,g_{\mu\nu}\left(  x\right)  $ are the $d$
dimensional shadows of their higher dimensional counterparts and the potential
is unique $V\left(  \phi\right)  =\lambda\phi^{\frac{2d}{d-2}}$ up to an
undetermined dimensionless constant $\lambda.$ The standard curvature term
with the Newton constant $G$ is \textit{not permitted} as a consequence of the
gauge symmetries of the parent theory. So there are no dimensionful constants
in this theory. In this shadow, due to the predetermined constant (fixed by
the gauge symmetries in 2T-gravity),
\begin{equation}
z_{d}\equiv\frac{d-2}{8\left(  d-1\right)  },
\end{equation}
there is an emergent local scaling (Weyl) symmetry in $d$ dimensions which is
a remnant of general coordinate transformations in the extra 1+1 dimensions
\cite{geometry 2T gravity}. Since the coefficient of $R\left(  g\right)  $
must be positive, the dilaton must have the wrong sign kinetic energy to
satisfy the Weyl symmetry, so $\phi$ is a ghost. Using the Weyl gauge symmetry
the shadow dilaton can be gauge fixed to a constant $\phi_{0}$ (thus
eliminating the ghost which would also have been a Goldstone boson after
condensation), yielding precisely Einstein's General Relativity with an
arbitrary cosmological constant $S_{grav}=\int d^{d}x\sqrt{-g}\{\phi_{0}%
^{2}R\left(  g\right)  -\lambda\phi_{0}^{2d/(d-2)}\},$ where the condensate
$\phi_{0}^{2}$ must be interpreted in terms of Newton's constant $\phi_{0}%
^{2}=\left(  16\pi G_{d}\right)  ^{-1}.$

Matter fields can be added to 2T-gravity, including Klein-Gordon type scalars,
Dirac or Weyl spinors and Yang-Mills type vectors, all in $d+2$ dimensions.
There are special restrictions on each one of these, on the form of their
kinetic energies, and the forms of permitted interactions among themselves and
with the gravitational fields$.$ These restrictions emerge from the underlying
gauge symmetry.

Here we are only concerned with the conformal shadow of this theory when it
includes only gravity coupled to scalar fields. These scalar fields are all
the elementary scalars that one would introduce in a complete theory (thus
including the dilaton, inflaton, Higgs, SUSY partners, GUT scalars, etc.). The
emerging conformal shadow then has the following form in the language of
ordinary field theory in $d$ dimensions with one time \cite{2T gravity}%
\cite{geometry 2T gravity}%
\begin{equation}
S=\int d^{d}x\sqrt{-g}\left(
\begin{array}
[c]{c}%
\frac{1}{2}g^{\mu\nu}\partial_{\mu}\phi\partial_{\nu}\phi-\frac{1}{2}g^{\mu
\nu}\sum_{i}\partial_{\mu}s_{i}\partial_{\nu}s_{i}\\
+z_{d}\left(  \phi^{2}-\sum_{i}s_{i}^{2}\right)  R\left(  g\right)
-\phi^{2d/(d-2)}f\left(  \frac{s_{i}}{\phi}\right)
\end{array}
\right)  \label{starting theory}%
\end{equation}
where $f\left(  s_{i}/\phi\right)  $ is an arbitrary function of its arguments
$s_{i}/\phi.$ This shadow automatically has the local Weyl scale symmetry,
\begin{equation}
g_{\mu\nu}^{\prime}=e^{2\lambda\left(  x\right)  }g_{\mu\nu},\;\phi^{\prime
}=e^{(1-d/2)\lambda\left(  x\right)  }\phi,\;s_{i}^{\prime}=e^{(1-d/2)\lambda
\left(  x\right)  }s_{i}, \label{Weyl transformation}%
\end{equation}
with an arbitrary gauge parameter $\lambda\left(  x\right)  ,$ which can be
verified directly in $d$ dimensions. Indeed the predicted special value of
$z_{d}$ and the form of the potential $V\left(  \phi,s_{i}\right)
=\phi^{2d/(d-2)}f\left(  \frac{s_{i}}{\phi}\right)  $ that scales like
$V\left(  t\phi,ts_{i}\right)  =t^{2d/(d-2)}V\left(  \phi,s_{i}\right)  $, are
crucial properties to realize this local symmetry. We emphasize that a Weyl
symmetry was not one of the gauge symmetries of the parent 2T-gravity field
theory in $d+2$ dimensions, rather it emerges in the conformal shadow in $d$
dimensions as a remnant from the general coordinate symmetry in the extra
dimensions \cite{geometry 2T gravity}. Thus the Weyl symmetry is a signature
of 2T physics.

In this coupling of gravity to matter, demanded by 2T-physics, there is an
interesting model independent physics prediction \cite{2T gravity} to be
emphasized. Every physical scalar $s_{i}$ in the complete field theory must be
a conformal scalar that couples to the curvature term just like the dilaton.
But to avoid being ghosts the $s_{i}$ must have the opposite relative sign in
the kinetic term. Then in the conformal shadow the curvature term is predicted
to take the form $(\phi^{2}-\sum_{i}s_{i}^{2})R\left(  g\right)  $ with a
required relative minus sign as shown. Hence the gravitational constant must
emerge from the condensates of all the scalars, not only the dilaton's. This
predicts a physical effect, that the effective gravitational constant
$G\sim(\phi^{2}-\sum_{i}s_{i}^{2})^{-1}$ is not really a constant, rather it
must increase after every phase transition of the universe as a whole. Since
the dominant part of each field is the condensate after the phase transition,
ignoring the small fluctuations, the effective gravitational constant is
approximately a constant in between the phase transitions. Thus the Newton
constant we measure today cannot be the same as the analogous constant before
the various transitions occurred, such as inflation, grand unification, SUSY
breaking, electroweak symmetry breaking, etc.. Of course the earlier phase
transitions supply the dominant condensates in the sum.

There is also the curious possibility that $G\sim(\phi^{2}-\sum_{i}s_{i}%
^{2})^{-1}$ could turn negative if the other scalars dominate over the dilaton
in some regions of the universe, or in the history of the universe, thus
producing antigravity in those parts of spacetime. The effects of this idea on
cosmology is one of the motivations that led us to investigate the solutions
of this theory, as we will do in the rest of this paper. Interestingly, we
found that the Big Bang is related to the vanishing of the gauge invariant
$(1-\sum_{i}s_{i}^{2}/\phi^{2}$) at which point the effective $G$ changes
sign. The familiar portion of the universe (with positive $G)$ and its
evolution is described by positive values of the factor $(1-\sum_{i}s_{i}%
^{2}/\phi^{2}$) starting with zero at the Big Bang and staying in the positive
region throughout the history of the universe. The value of the gauge
invariant quantity $(1-\sum_{i}s_{i}^{2}/\phi^{2})$$\;$oscillates in the
positive region before it reaches its asymptotic value 1. As a result, the
vanishing of this \textit{gauge invariant} quantity determines the Big
Bang.\footnote{Meanwhile, the behavior of the gauge dependent quantity
$G\sim(\phi^{2}-\sum_{i}s_{i}^{2})^{-1}$ varies according to the gauge choice;
for example it can be taken to be a positive constant in the Einstein frame,
but only in spacetime regions where it is positive.}

\subsection{The model for the Big Bang and inflation}

To solve the differential equations for inflation $\left(  \ref{00}\right)
,\left(  \ref{11}\right)  ,\left(  \ref{eom sigma}\right)  ,$ and even harder
ones, we will start with a theory of the type (\ref{starting theory}) that
includes two scalar fields $\phi,s,$ with a Weyl local scaling symmetry. Our
strategy is to use the Weyl symmetry to make some convenient gauge choices. In
one gauge the theory will reduce to the standard inflaton theory in
Eqs.(\ref{inflaton potential},\ref{theory}), while in another gauge it will
reduce to a completely solvable theory. Since each one is a gauge choice, the
solvable theory is dual to the inflaton theory\footnote{This duality, which is
in the form of a familiar gauge transformation in the present case, is a
simple example of a rich set of dualities among 1T physical systems predicted
by 2T-physics, but missed systematically in 1T-physics. Choosing a Weyl gauge
in the present context amounts to choosing how to embed $d$ dimensions in
$d+2$ dimensions, thus creating a perspective of how an observer in $d$
dimensions perceives some \textquotedblleft shadow\textquotedblright\ of
phenomena that occur in $d+2$ dimensions. For a recent summary of 2T-physics
that includes a description of such phenomena see \cite{GMtalk}.}. The
inflaton equations $\left(  \ref{00}\right)  ,\left(  \ref{11}\right)
,\left(  \ref{eom sigma}\right)  $ are then solved by transforming the
solution from the fully solvable version.

The starting point is then the action%
\begin{equation}
S=\int d^{4}x\sqrt{-g}\left(  \frac{1}{2}g^{\mu\nu}\partial_{\mu}\phi
\partial_{\nu}\phi-\frac{1}{2}g^{\mu\nu}\partial_{\mu}s\partial_{\nu}%
s+\frac{1}{12}\left(  \phi^{2}-s^{2}\right)  R-\phi^{4}f\left(  \frac{s}{\phi
}\right)  \right)  , \label{action}%
\end{equation}
where we have used $z_{4}=\frac{1}{12}.$ Here $f$ can be an arbitrary function
of $\frac{s}{\phi}$, but will be later set to $f\left(  \frac{s}{\phi}\right)
=c\left(  \frac{s}{\phi}\right)  ^{4}+b$ in the present application to
reproduce the model of Eq.(\ref{inflaton potential})$.$ Here $\phi$ and $s $
are real scalars while $\phi$ has the wrong sign in the kinetic term. This
makes $\phi$ a ghost degree of freedom. With the special coefficient $\frac
{1}{12}$ in the coupling of the scalars fields to the Ricci curvature, this
action is invariant under the local Weyl transformation in Eq.$\left(
\ref{Weyl transformation}\right)  $. Due to this local Weyl symmetry, we can
eliminate the ghost degree of freedom by gauge fixing. So the theory is
actually \emph{ghost free}. This Weyl symmetry will play a crucial roll in
solving the theory $\left(  \ref{theory}\right)  .$ Note that under the Weyl
transformations the quantity $\left(  1-s^{2}/\phi^{2}\right)  $ is gauge
invariant, so the sign or the zeros of the effective gravitational coupling
$\left(  \phi^{2}-s^{2}\right)  \sim G^{-1}$ are the same for all gauge choices.

Varying the action $\left(  \ref{action}\right)  $ with respect to all its
degrees of freedom, we derive the equations of motion%
\begin{gather}
R_{\mu\nu}-\frac{1}{2}Rg_{\mu\nu}=T_{\mu\nu}\label{Einstein}\\
\nabla^{2}\phi=\frac{1}{6}\phi R-4\phi^{3}f\left(  \frac{s}{\phi}\right)
+\frac{s}{\phi^{2}}f^{\prime}\left(  \frac{s}{\phi}\right) \label{eom phi}\\
\nabla^{2}s=\frac{1}{6}sR+\phi^{3}f^{\prime}\left(  \frac{s}{\phi}\right)
\label{eom s}%
\end{gather}
where $f^{\prime}\left(  \frac{s}{\phi}\right)  $ denotes the derivative of
$f$ with respect to its argument $\frac{s}{\phi}.$ The energy momentum tensor
$T_{\mu\nu}$ is%
\begin{equation}
T_{\mu\nu}=\frac{6}{\left(  \phi^{2}-s^{2}\right)  }\left[
\begin{array}
[c]{c}%
-\partial_{\mu}\phi\partial_{\nu}\phi+\partial_{\mu}s\partial_{\nu}s-\frac
{1}{6}\left(  g_{\mu\nu}\nabla^{2}-\nabla_{\mu}\partial_{\nu}\right)  \left(
\phi^{2}-s^{2}\right) \\
+g_{\mu\nu}\left(  \frac{1}{2}\partial\phi\cdot\partial\phi-\frac{1}%
{2}\partial s\cdot\partial s-\phi^{4}f\left(  \frac{s}{\phi}\right)  \right)
\end{array}
\right]  \label{energy momentum tensor}%
\end{equation}
The above equations are valid for all gauge choices of the Weyl symmetry.

Now we select gauges. To start with, we will choose the so called Einstein
gauge where the coefficient in front of the Ricci curvature is set to the
usual gravitational constant%
\begin{equation}
\frac{1}{12}\left(  \phi_{E}^{2}-s_{E}^{2}\right)  =\frac{1}{2\kappa^{2}}
\label{EinsteinGauge}%
\end{equation}
The subscripts $E$ indicate that these fields, including the metric $\left(
g_{E}\right)  _{\mu\nu},$ are in the Einstein gauge. So in this gauge,
$\phi_{E}$ is related to the field $s_{E\text{ }}$ as
\begin{equation}
\phi_{E}=\pm\left(  s_{E}^{2}+\frac{6}{\kappa^{2}}\right)  ^{\frac{1}{2}}.
\label{phiS}%
\end{equation}
The $\pm$ plays no role because $\phi_{E}$ appears always quadratically. Now,
inserting this into the action $\left(  \ref{starting theory}\right)  ,$ and
further using the following field redefinition
\begin{equation}
s_{E}=\frac{\sqrt{6}}{\kappa}\sinh\left(  \frac{\kappa\sigma}{\sqrt{6}%
}\right)  , \label{ssigma}%
\end{equation}
we find that the gauge fixed form of the action $\left(  \ref{action}\right)
$ reduces to the inflaton action $\left(  \ref{theory}\right)  $ and relates
the general inflaton potential $V\left(  \sigma\right)  $ to the general
function $f\left(  s/\phi\right)  .$ This result indicates that the physics of
$\left(  \ref{theory}\right)  $ is completely equivalent to the physics of
$\left(  \ref{action}\right)  $ for any desired potential. Hence we can use
the Weyl symmetric version $\left(  \ref{action}\right)  $ in other gauges to
tackle the solution of the inflaton theory.

Now suppose we wish to discuss the case of a conformally flat spacetime
defined by the line element%
\begin{equation}
ds^{2}=a^{2}\left(  x\right)  \left(  \eta_{\mu\nu}dx^{\mu}dx^{\nu}\right)
\end{equation}
where $\eta_{\mu\nu}$ is the flat Minkowski metric while the scale factor
$a\left(  x\right)  $ is an \textit{arbitrary} function of spacetime. The
Robertson-Walker metric (\ref{FRW}) that we discuss later is a special case
where the scale factor is a function of only time, but for now we are
considering the more general spacetime dependence in $a\left(  x\right)  .$

The equations above (\ref{Einstein}-\ref{eom s}) simplify for the conformally
flat metric. They now contain three fields $a\left(  x\right)  ,\phi\left(
x\right)  $ and $s\left(  x\right)  .$ Under the Weyl transformations we can
form two invariants, which we can choose as (here we state the results more
generally for any $d,$ but for the application we use $d=4$)
\begin{equation}
\tilde{\phi}=a^{(d/2-1)}\phi\text{ and }\tilde{s}=a^{(d/2-1)}s.
\end{equation}
All the equations above, when specialized to the conformal spacetime, can be
written in terms of only these two gauge invariants.

To proceed, we will choose a Weyl gauge where the metric is actually flat. In
this gauge we have $a_{flat}\left(  x\right)  =1,$ while $\phi_{flat}%
,s_{flat}$ are still arbitrary but they are equal to the gauge invariants
defined above%
\begin{equation}
a_{flat}\left(  x\right)  =1:\;\tilde{\phi}=\phi_{flat},\;\tilde{s}=s_{flat}.
\end{equation}
This shows that by choosing the \textquotedblleft flat gauge\textquotedblright%
\ we can extract all the gauge invariant information in the equations of
motion (\ref{Einstein}-\ref{eom s}) for a conformally flat spacetime.

In the flat gauge, since $R\left(  \eta\right)  =0$ and $\sqrt{-\eta}=1,$ the
action $\left(  \ref{action}\right)  $ becomes
\begin{equation}
S=\int d^{d}x\left(  \frac{1}{2}\left(  \partial_{\mu}\phi_{flat}\right)
^{2}-\frac{1}{2}\left(  \partial_{\mu}s_{flat}\right)  ^{2}-\phi
_{flat}^{2d/(d-2)}f\left(  \frac{s_{flat}}{\phi_{flat}}\right)  \right)
\end{equation}
where, the conformally flat metric $g_{\mu\nu}$ is actually flat $\eta_{\mu
\nu}.$

Now we note that in this gauge the two scalar fields $\phi_{flat}$ and
$s_{flat}$ (equivalently the two gauge invariants $\tilde{\phi},\tilde{s}$)
will have completely decoupled dynamics if $f\left(  s/\phi\right)  $ takes
the special form $f\left(  s/\phi\right)  =c\left(  s/\phi\right)
^{d/(d-2)}+b.$ This is when the inflaton potential takes the special form in
Eq.(\ref{inflaton potential}) when $d=4.$ In this case the action describes
two completely decoupled scalars $\phi_{flat}$ and $s_{flat}$ were
$\phi_{flat}$ has a wrong sign for its kinetic term%
\begin{equation}
S=\int d^{d}x\left(  \frac{1}{2}\left(  \partial_{\mu}\phi_{flat}\right)
^{2}-\frac{1}{2}\left(  \partial_{\mu}s_{flat}\right)  ^{2}-b\phi
_{flat}^{2d/(d-2)}-cs_{flat}^{2d/(d-2)}\right)  . \label{action in flat gauge}%
\end{equation}
The equations of motion (\ref{eom phi},\ref{eom s}) in this gauge are%
\begin{equation}
\square\phi_{flat}=-\frac{2d}{d-2}b\phi_{flat}^{(d+2)/(d-2)},\;\;\square
s_{flat}=\frac{2d}{d-2}cs_{flat}^{(d+2)/(d-2)}. \label{decoupled}%
\end{equation}
where $\square=\eta^{\mu\nu}\partial_{\mu}\partial_{\nu}.$ In addition to
these equations one should also impose the constraints that follow from
general coordinate symmetry (which has been gauge fixed) that imply, through
Eq.(\ref{Einstein}), that the stress tensor in
Eq.(\ref{energy momentum tensor}) specialized to the flat gauge must also
vanish
\begin{equation}
T_{\mu\nu}^{flat}=0. \label{T00}%
\end{equation}
There remains solving the equations in this gauge. It should be emphasized
that the decoupling of $\phi_{flat}$ and $s_{flat}$ is valid for the general
conformally flat metric, that is, for any spacetime dependence of $a\left(
x\right)  ,$ and therefore for any $a_{E}\left(  x\right)  $ in the Einstein
gauge in the standard form of the theory of Eqs.(\ref{inflaton potential}%
,\ref{theory}) (i.e. not only the Robertson-Walker case).

In solving these decoupled equations we should not forget that the acceptable
solutions in the flat gauge must still satisfy the gauge invariant requirement
that the sign and the zeros of $\left(  1-s^{2}/\phi^{2}\right)  $ must be the
same in all gauge fixed versions. In particular if we wish to relate to the
Einstein gauge in which $\left(  1-s_{E}^{2}/\phi_{E}^{2}\right)
=(2z_{d}\kappa^{2}\phi_{E}^{2})^{-1}$ is always positive (see Eq.(\ref{phiS}%
)), then the corresponding acceptable solutions in the flat gauge must also
satisfy $\left(  1-s_{flat}^{2}/\phi_{flat}^{2}\right)  \geq0.$ The zero in
the flat gauge at $s_{flat}/\phi_{flat}=\pm1$ may be attained only when in the
Einstein gauge $s_{E}$ and $\phi_{E}$ tend to infinity according to their
relation in Eq.(\ref{phiS}). According to the relation to the inflaton
$\sigma$ in Eq.(\ref{ssigma}) the zero can happen in any gauge only when the
inflaton $\sigma$ in the Einstein frame blows up. We will see through exact
solutions that this gauge invariant zero corresponds to the Big Bang!

\subsection{Relating the Einstein and flat gauges}

Notice that the quantity $\frac{s}{\phi}$ is scale invariant, so it can be
expressed in various Weyl gauges as follows%
\begin{equation}
\frac{s}{\phi}=\frac{s_{flat}}{\phi_{flat}}=\frac{s_{E}}{\phi_{E}},
\end{equation}
and by using Eq.(\ref{ssigma}) this allows us to relate $\sigma$ to
$s_{flat}/\phi_{flat}$ as follows
\begin{equation}
\frac{s_{flat}}{\phi_{flat}}=s_{E}\left(  s_{E}^{2}+\frac{1}{2z_{d}\kappa^{2}%
}\right)  ^{-\frac{1}{2}}=\tanh\left(  \sqrt{2z_{d}}\kappa\sigma\right)  .
\label{relate y to sigma}%
\end{equation}
Thus the inflaton field $\sigma$ in the Einstein gauge is related to the flat
gauge fields as follows%
\begin{equation}
\sigma=\frac{1}{\sqrt{8z_{d}}\kappa}\ln\left\vert \frac{\phi_{flat}+s_{flat}%
}{\phi_{flat}-s_{flat}}\right\vert \underset{d=4}{\rightarrow}\frac{\sqrt{6}%
}{2\kappa}\ln\left\vert \frac{\phi_{flat}+s_{flat}}{\phi_{flat}-s_{flat}%
}\right\vert \label{transform AW to sigma}%
\end{equation}

The other field in the Einstein gauge is the scale factor $a_{E}\left(
\tau\right)  $ which appears in the equations (\ref{00}-\ref{eom sigma}) we
wish to solve. It is related to the flat fields by a Weyl gauge transformation
involving a local parameter $\lambda\left(  x\right)  $ consistent with
Eq.$\left(  \ref{Weyl transformation}\right)  $. Hence we can determine
$\lambda\left(  x\right)  $ through the equation%
\begin{equation}
s_{E}=e^{(1-d/2)\lambda\left(  x\right)  }s_{flat},\text{ \ }\phi
_{E}=e^{(1-d/2)\lambda\left(  x\right)  }\phi_{flat}=\left(  s_{E}^{2}%
+\frac{1}{2z_{d}\kappa^{2}}\right)  ^{\frac{1}{2}}
\label{duality transformation s}%
\end{equation}
So, consistent with the Einstein gauge of Eq.(\ref{EinsteinGauge}), we can
write
\begin{equation}
e^{(d-2)\lambda\left(  x\right)  }=2z_{d}\kappa^{2}\left(  \phi_{flat}%
^{2}-s_{flat}^{2}\right)  ,\text{ for }\left(  \phi_{flat}^{2}-s_{flat}%
^{2}\right)  \geq0. \label{Weyl transform}%
\end{equation}
Now, using Eq.$\left(  \ref{Weyl transform}\right)  $ we can figure out the
scale factor in the Einstein gauge as%
\begin{equation}
a_{E}=e^{\lambda\left(  x\right)  }a_{flat}=\left[  2z_{d}\kappa^{2}\left(
\phi_{flat}^{2}-s_{flat}^{2}\right)  \right]  ^{1/(d-2)}\underset
{d=4}{\rightarrow}\frac{\kappa}{\sqrt{6}}\left(  \phi_{flat}^{2}-s_{flat}%
^{2}\right)  ^{1/2} \label{transform AW to AE}%
\end{equation}
where we used $a_{flat}=1.$ So, Eqs.(\ref{transform AW to sigma}%
,\ref{transform AW to AE}) provide the duality transformation for relating the
Einstein-gauge fields $\left(  a_{E}\left(  x\right)  ,\sigma\left(  x\right)
\right)  $ to the flat-gauge fields $\left(  \phi_{flat}\left(  x\right)
,s_{flat}\left(  x\right)  \right)  .$ In these expressions we may substitute
the gauge invariants $\tilde{\phi},\tilde{s}$ instead of $\phi_{flat}%
,s_{flat}.$

It should be emphasized that this approach works in any dimension, not just
four. By solving the simpler decoupled equations (\ref{decoupled}) for
$\left(  \phi_{flat}\left(  x\right)  ,s_{flat}\left(  x\right)  \right)  ,$
and then using this duality (which depends on $d$), we will obtain the
solutions for the much more complicated coupled differential equations for the
fields $\left(  a_{E}\left(  x\right)  ,\sigma\left(  x\right)  \right)  $ for
any conformally flat metric in the Einstein gauge, i.e. any $g_{\mu\nu}%
^{E}=a_{E}\left(  x\right)  \eta_{\mu\nu}$ with arbitrary $x^{\mu}$
dependence. Special examples of such spacetime metrics include $AdS_{d},$
$AdS_{d-1}\times S^{1},$ $\cdots$, $AdS_{d-n}\times S^{n},$ any maximally
symmetric space, any conformally flat space including singular ones, etc. ,
and of course the Robertson-Walker expanding universe that we wish to discuss
next. It is worth noticing that these special conformally flat spacetimes in
$d$ dimensions, which can all be mapped to our decoupled system, have a hidden
global $SO(d,2)$ symmetry, since they are obtained as shadows of the
completely flat spacetime in $d+2$ dimensions in the context of 2T-physics
\cite{Dual field theories}\cite{Dual fild theories fermion}.

\subsection{Solution of inflation equations}

For the problem of inflation we need to specialize to the much simpler
homogeneous fields that depend only on the conformal time $\tau$ and take
$d=4.$ The solution for the complicated equations (\ref{00}-\ref{eom sigma})
for $a_{E}\left(  \tau\right)  $ and $\sigma\left(  \tau\right)  $ can now be
obtained by solving the decoupled simple equations $\left(  \phi_{flat}\left(
\tau\right)  ,s_{flat}\left(  \tau\right)  \right)  $ plus our duality
transformation (\ref{transform AW to sigma},\ref{transform AW to AE}). \ Then,
equations (\ref{decoupled}) take the form%
\begin{equation}
-\ddot{\phi}_{flat}=-4b\phi_{flat}^{3},\;\;-\ddot{s}_{flat}=4cs_{flat}^{3}.
\label{elliptic}%
\end{equation}
where the derivatives are with respect to $\tau.$ A first integral is obtained
in the form
\begin{equation}
-\frac{\dot{\phi}_{flat}^{2}}{2}+b\phi_{flat}^{4}=-E_{\phi},\text{ \ \ }%
\frac{\dot{s}_{flat}^{2}}{2}+cs_{flat}^{4}=E_{s} \label{eom phi flat}%
\end{equation}
Using the $00$ component of Eq.$\left(  \ref{Einstein}\right)  $, or
Eq.(\ref{T00}), we find that the constants are related $E_{s}=E_{\phi}=E,$
where $E$ is the energy density of $s_{flat}$ and $\left(  -E\right)  $ is the
energy density of $\phi_{flat},$ so%
\begin{equation}
\frac{\dot{\phi}_{flat}^{2}}{2}-b\phi_{flat}^{4}=\frac{\dot{s}_{flat}^{2}}%
{2}+cs_{flat}^{4}=E. \label{Einstein 00}%
\end{equation}
This implies that the total energy density in the flat gauge is zero which
makes perfect sense. If the total energy density were not zero then we could
not have a static flat metric.

The nature of the solutions can now be ascertained intuitively because
(\ref{Einstein 00}) looks like the problem of a non-relativistic particle
moving in a quartic potential (upside or downside, depending on the signs of
$b,c$) at a fixed energy $E$ (which can be positive, negative or zero).

More precisely, equations (\ref{elliptic}) are second order nonlinear
differential equations. A first integral is already given by
Eq.(\ref{Einstein 00}), and this leads to first order non-linear differential
equations, $\dot{\phi}_{flat}=\pm\sqrt{2E+2b\phi_{flat}^{4}}~$and $\dot
{s}_{flat}=\pm\sqrt{2E-2cs_{flat}^{4}}$, that are easily integrated. For the
case $E=0$ the solutions are simple. For $E\neq0$ the solutions are expressed
in terms of the famous Jacobi elliptic function $cn\left(  z|\frac{1}%
{2}\right)  $ as seen below. The function $cn\left(  z|m\right)  $ with a more
general label $m$ is a doubly periodic meromorphic function (similar to the
cosine function $\cos\left(  \omega z\right)  $) where $m$ is the parameter
which determines the period of the function (for more information see
\cite{Hand book} and appendix \ref{Jacobi elliptic functions})). All the
possible solutions for different regions of the parameters are listed below,
where we have abbreviated $cn\left(  z\right)  \equiv cn\left(  z|\frac{1}%
{2}\right)  $ and defined the parameters
\begin{equation}
\;\zeta\equiv\left\vert \frac{4b}{c}\right\vert ^{\frac{1}{4}},\text{~~}%
\widetilde{\tau}\equiv2\left\vert cE\right\vert ^{\frac{1}{4}}\tau
,~\ \delta=\text{constant.}%
\end{equation}
Here $\zeta$ is a dimensionless parameter that depends on the ratio of $b$ and
$c,$ $\widetilde{\tau}$ is a dimensionless conformal time. The purpose of
introducing $\zeta$ and $\widetilde{\tau}$ is just to simplify the expressions
in the tables below$.$ However, we will still use the dimensionful $\tau$ in
some of the solutions where $c$ or $E$ or both of them are zero. The constant
$\delta$ is a relative shift in dimensionless conformal time. The origin of
the conformal time can be changed arbitrarily because of translation symmetry
of the equations under $\tau\rightarrow\tau+\tau_{0}.$ Therefore $\tau_{0}$
has no physical meaning but the relative time $\delta$ which corresponds to
initial conditions has physical meaning.

There are four free parameters in the solutions of these two decoupled second
order differential equations, which are $E_{\phi},E_{s},\delta$ and $\tau
_{0}.$ We have just argued that $\tau_{0}$ is not a physical quantity that can
be fixed arbitrarily, also Eq.$\left(  \ref{Einstein}\right)  $ requires
$E_{\phi}=E_{s}=E$. This reduces the number of undetermined integration
parameters associated with boundary conditions to only the two parameters
$E,\delta$. Together with the two parameters $b$ and $c$ that determine the
shape of the inflaton potential, we have totally four parameters in the theory
which can be adjusted to fit phenomenological observations.

The following three tables elaborate on all possible solutions of
Eqs.(\ref{elliptic})$,$ corresponding to all possible solutions of the
Einstein equations (\ref{00}-\ref{eom sigma}) in the Einstein frame%

\begin{gather*}%
\begin{tabular}
[c]{|c|c|c|c|c|}\hline
$b$ & $E$ & $c$ & $\phi_{flat}$ & $s_{flat}$\\\hline
$b>0$ & $E>0$ & $c>0$ & $\left(  \frac{E}{b}\right)  ^{\frac{1}{4}}\left[
\frac{1-cn\left(  \zeta\widetilde{\tau}\right)  }{1+cn\left(  \zeta
\widetilde{\tau}\right)  }\right]  ^{\frac{1}{2}}$ & $\left(  \frac{E}%
{c}\right)  ^{\frac{1}{4}}cn\left(  \widetilde{\tau}+\delta\right)  $\\\hline
&  & $c=0$ & $\left(  \frac{E}{b}\right)  ^{\frac{1}{4}}\left[  \frac
{1-cn\left(  \zeta\widetilde{\tau}\right)  }{1+cn\left(  \zeta\widetilde{\tau
}\right)  }\right]  ^{\frac{1}{2}}$ & $\sqrt{2E}\left(  \tau+E^{\frac{-1}{4}%
}\delta\right)  $\\\hline
&  & $c<0$ & $\left(  \frac{E}{b}\right)  ^{\frac{1}{4}}\left[  \frac
{1-cn\left(  \zeta\widetilde{\tau}\right)  }{1+cn\left(  \zeta\widetilde{\tau
}\right)  }\right]  ^{\frac{1}{2}}$ & $\left\vert \frac{E}{c}\right\vert
^{\frac{1}{4}}\left[  \frac{1-cn\left(  \sqrt{2}\widetilde{\tau}%
+\delta\right)  }{1+cn\left(  \sqrt{2}\widetilde{\tau}+\delta\right)
}\right]  ^{\frac{1}{2}}$\\\hline
& $E=0$ & $c>0$ & $\frac{1}{\pm\sqrt{2b}\tau}$ & no solution\\\hline
&  & $c=0$ & $\frac{1}{\pm\sqrt{2b}\tau}$ & $s_{0}=const$\\\hline
&  & $c<0$ & $\frac{1}{\pm\sqrt{2b}\tau}$ & $\frac{1}{\pm\sqrt{-2c}\tau
+\kappa\delta}$\\\hline
& $E<0$ & $c\eqslantgtr0$ & $\left\vert \frac{E}{b}\right\vert ^{\frac{1}{4}%
}\frac{1}{cn\left(  \frac{\zeta}{\sqrt{2}}\widetilde{\tau}\right)  }$ & no
solution\\\hline
&  & $c<0$ & $\left\vert \frac{E}{b}\right\vert ^{\frac{1}{4}}\frac
{1}{cn\left(  \frac{\zeta}{\sqrt{2}}\widetilde{\tau}\right)  }$ & $\left\vert
\frac{E}{c}\right\vert ^{\frac{1}{4}}\frac{1}{cn\left(  \widetilde{\tau
}+\delta\right)  }$\\\hline
\end{tabular}
\\
\text{Table 1 - Solutions for }b>0.\;
\end{gather*}

\bigskip%

\begin{gather*}%
\begin{tabular}
[c]{|c|c|c|c|c|}\hline
$b$ & $E$ & $c$ & $\phi_{flat}$ & $s_{flat}$\\\hline
$b=0$ & $E>0$ & $c>0$ & $\pm\sqrt{2E}\left(  \tau+E^{\frac{-1}{4}}%
\delta\right)  $ & $\left(  \frac{E}{c}\right)  ^{\frac{1}{4}}cn\left(
\widetilde{\tau}\right)  $\\\hline
&  & $c=0$ & $\pm\sqrt{2E}\left(  \tau+E^{\frac{-1}{4}}\delta\right)  $ &
$\sqrt{E}\tau$\\\hline
&  & $c<0$ & $\pm\sqrt{2E}\left(  \tau+E^{\frac{-1}{4}}\delta\right)  $ &
$\frac{\zeta}{\sqrt{2}}\left\vert \frac{E}{b}\right\vert ^{\frac{1}{4}}\left[
\frac{1-cn\left(  \widetilde{\tau}\right)  }{1+cn\left(  \widetilde{\tau
}\right)  }\right]  ^{\frac{1}{2}}$\\\hline
& $E=0$ & $c>0$ & $\phi_{0}=const$ & no solution\\\hline
&  & $c=0$ & $\phi_{0}=const$ & $s_{0}=const$\\\hline
&  & $c<0$ & $\phi_{0}=const$ & $\frac{1}{\pm\sqrt{-2c}\tau+\kappa\delta}%
$\\\hline
& $E<0$ & all $c$ & no solution & \\\hline
\end{tabular}
\\
\text{Table 2 - Solutions for }b=0.\;
\end{gather*}

\bigskip%

\begin{gather*}%
\begin{tabular}
[c]{|c|c|c|c|c|}\hline
$b$ & $E$ & $c$ & $\phi_{flat}$ & $s_{flat}$\\\hline
$b<0$ & $E>0$ & $c>0$ & $\left\vert \frac{E}{b}\right\vert ^{\frac{1}{4}%
}cn\left(  \frac{\zeta}{\sqrt{2}}\widetilde{\tau}\right)  $ & $\left\vert
\frac{E}{c}\right\vert ^{\frac{1}{4}}cn\left(  \widetilde{\tau}+\delta\right)
$\\\hline
&  & $c=0$ & $\left\vert \frac{E}{b}\right\vert ^{\frac{1}{4}}cn\left(
\frac{\zeta}{\sqrt{2}}\widetilde{\tau}\right)  $ & $\pm\sqrt{2E}\left(
\tau+E^{\frac{-1}{4}}\delta\right)  $\\\hline
&  & $c<0$ & $\left\vert \frac{E}{b}\right\vert ^{\frac{1}{4}}cn\left(
\frac{\zeta}{\sqrt{2}}\widetilde{\tau}\right)  $ & $\left\vert \frac{E}%
{c}\right\vert ^{\frac{1}{4}}\left[  \frac{1-cn\left(  \widetilde{\tau}%
+\delta\right)  }{1+cn\left(  \widetilde{\tau}+\delta\right)  }\right]
^{\frac{1}{2}}$\\\hline
& $E\leq0$ & all $c$ & no solution & \\\hline
\end{tabular}
\\
\text{Table 3 - Solutions for }b<0.\;
\end{gather*}
Since all solutions of $\phi_{flat}$ and $s_{flat}$ are known, it is
straightforward to calculate $\sigma$ and $a_{E}$ from the duality (or gauge)
transformation in the previous subsection and the results listed in Tables
1,2,3. One can check that Eq.$\left(  \ref{transform AW to AE}\right)  $ and
Eq.$\left(  \ref{transform AW to sigma}\right)  $ are actually solutions to
differential equations $\left(  \ref{00}\right)  ,\left(  \ref{11}\right)  $
and $\left(  \ref{eom sigma}\right)  $ by direct substitution provided
$\phi_{flat}$ and $s_{flat}$ are one of the possible solutions in Tables 1,2,3
that satisfies Eq.$\left(  \ref{Einstein 00}\right)  .$ Verifying this
requires only the second derivatives supplied by Eq.(\ref{elliptic}) and the
first derivatives supplied by Eq.(\ref{Einstein 00}). Of course, these are
consistent with the properties of the Jacobi elliptic functions listed in the appendix.

We have used conformal time as the evolution parameter, however, cosmologists
usually discuss time using physical time $t$ that is measured by a comoving
observer. To convert the conformal time to the physical time, one needs to
perform an integral,%

\begin{equation}
t=\int dt=\int a_{E}\left(  \tau\right)  d\tau. \label{physical time}%
\end{equation}
This follows from the definition of conformal time. In the example we are
going to discuss, the physical time will diverge logarithmically as conformal
time approaches a critical finite value $\tau_{\infty}.$ This is because as
conformal time approaches this critical value, our solution indicates that the
scale factor in the Einstein gauge diverges as $a_{E}\left(  \tau\right)
\sim\frac{1}{\tau-\tau_{\infty}}.$ With all the information given in this
section, we can proceed to build a phenomenological inflation model.

\section{A Model of Inflation}

\label{toy model}

In this section, we will construct a model of inflation that matches all of
the phenomenological constraints. The specific solution we are using is the
one in Table $1$ with $b>0,c>0,E>0.$ The solution for this range of parameter
space is%
\begin{equation}
\phi_{flat}=\left(  \frac{E}{b}\right)  ^{\frac{1}{4}}\left[  \frac
{1-cn\left(  \zeta\widetilde{\tau}\right)  }{1+cn\left(  \zeta\widetilde{\tau
}\right)  }\right]  ^{\frac{1}{2}},\;\;s_{flat}=\left(  \frac{E}{b}\right)
^{\frac{1}{4}}\frac{\xi}{\sqrt{2}}cn\left(  \widetilde{\tau}+\delta\right)
\label{flat}%
\end{equation}
From Eqs.(\ref{transform AW to AE}) and (\ref{transform AW to sigma}) the
scale factor in the Einstein gauge and the inflaton can be written as (recall
$\zeta\equiv\left\vert \frac{4b}{c}\right\vert ^{\frac{1}{4}}$ and
$\widetilde{\tau}\equiv2\left\vert cE\right\vert ^{\frac{1}{4}}\tau$)%
\begin{equation}
a_{E}=\frac{\kappa}{\sqrt{6}}\left(  \frac{E}{b}\right)  ^{\frac{1}{4}%
}\left\{  \left[  \frac{1-cn\left(  \zeta\widetilde{\tau}\right)
}{1+cn\left(  \zeta\widetilde{\tau}\right)  }\right]  -\frac{\zeta^{2}}%
{2}\left[  cn\left(  \widetilde{\tau}+\delta\right)  \right]  ^{2}\right\}
^{\frac{1}{2}} \label{AE}%
\end{equation}
and%
\begin{equation}
\sigma_{E}=\frac{\sqrt{6}}{\kappa}\frac{1}{2}\ln\left(  \frac{1+\frac{\zeta
}{\sqrt{2}}cn\left(  \widetilde{\tau}+\delta\right)  \left[  \frac{1+cn\left(
\zeta\widetilde{\tau}\right)  }{1-cn\left(  \zeta\widetilde{\tau}\right)
}\right]  ^{\frac{1}{2}}}{1-\frac{\zeta}{\sqrt{2}}cn\left(  \widetilde{\tau
}+\delta\right)  \left[  \frac{1+cn\left(  \zeta\widetilde{\tau}\right)
}{1-cn\left(  \zeta\widetilde{\tau}\right)  }\right]  ^{\frac{1}{2}}}\right)
\label{sigma as function of z and t}%
\end{equation}
It should be noted that the energy parameter $E$ determines the scale of
$a_{E},$ so $E$ could be fixed if we want to normalize $a_{E}$ in the
conventional way, $a_{E}\left(  today\right)  =1,$ but since we are discussing
the early universe we will keep it as shown above.

In the remainder of the discussion we will further specialize to a specific
value of the available constants by taking $\delta=0$ and $\zeta=\frac{1}{2}$
(or $64b=c)$. Although there is no reason \textit{\`{a} priori} to prefer this
model, we will take this as an example to show that our solution is compatible
with a phenomenological model of inflation. A study of phenomenologically
consistent more general parameter space, determined by numerical analysis is
expected in following papers.

The scale factor $a_{E}$ as function of $\widetilde{\tau}$ is plotted in Fig.1%

\begin{center}
\includegraphics[
height=2.6576in,
width=2.6576in
]%
{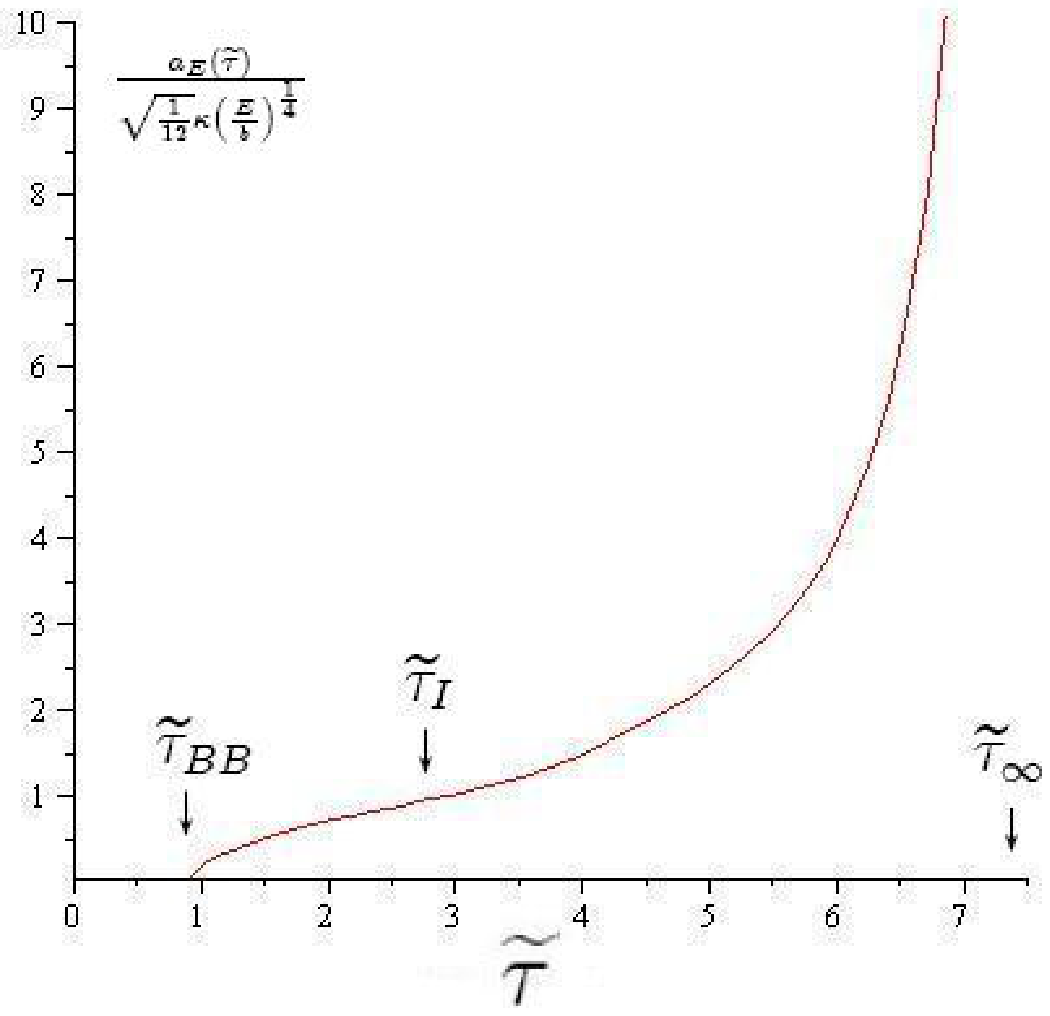}%
\\
Fig.1 - The scale factor $a_{E}\left(  \tilde{\tau}\right)  $ in the Einstein
frame, from Eq.$\left(  \ref{AE}\right)  $ with $\zeta=\frac{1}{2}$and
$\delta=0.$%
\label{1}%
\end{center}
The interesting behavior of this particular solution is that at $\widetilde
{\tau}=\widetilde{\tau}_{BB}\approx0.92$ the scale factor is exactly zero
$a_{E}\left(  \tau_{BB}\right)  =0$. This defines the origin for comoving time
$t\left(  \tau_{BB}\right)  =0$ at the Big Bang\footnote{{}At the moment of
the Big Bang $\tilde{\tau}_{BB},$ it also happens that the gauge invariant
variable ($1-s^{2}/\phi^{2}$) vanishes. The solution in terms of
($1-s_{flat}^{2}/\phi_{flat}^{2})$ is allowed to change sign in the flat
gauge, but if we insist that the physics must be the physics described in the
Einstein gauge in terms of the comoving time interval $0\leq t<\infty,$ then
we must confine the solution to only ($1-s_{flat}^{2}/\phi_{flat}^{2})\geq0.$
But it is interesting to note that in a more general gauge there are solutions
in which the gauge invariant quantity $\left(  1-s^{2}/\phi^{2}\right)  $ can
go through zero and change sign, thus making a transition to a region with
antigravity where the effective Newton constant $G\sim\left(  \phi^{2}%
-s^{2}\right)  ^{-1}$ is negative. Of course, applying a classical theory at
the Big Bang, where $\left(  1-s^{2}/\phi^{2}\right)  $ changes sign is
incomplete, however, this can be a guideline for a more complete quantum
theory of the Big Bang, with possibly new physics insights.}. Unlike pure
exponential inflation, where the conformal time of the Big Bang is equal to
$-\infty,$ our model gives a finite conformal time for the Big Bang. Of course
the finite value $\widetilde{\tau}_{BB}\approx0.92$ is not physically
significant since $\tau$ can be translated by an arbitrary amount as remarked
earlier. A finite conformal time for the Big Bang is a general property of our
scalar-tensor theory $\left(  \ref{theory}\right)  .$

After the Big Bang, the scale factor increases monotonically to infinity at
$\widetilde{\tau}_{\infty}\approx7.4$. Converting conformal time to physical
time $t$ for a comoving observer using Eq.$\left(  \ref{physical time}\right)
,$ the corresponding physical time is infinite\footnote{The area under this
curve up to some point $\widetilde{\tau}$ gives the comoving time $t$ (see
Eq.(\ref{physical time})). So, it is intuitive to define $t=0$ to correspond
to $\widetilde{\tau}_{BB}$ while $t=$infinity corresponds to $\widetilde{\tau
}=\widetilde{\tau}_{\infty}.$ The solution (\ref{AE}) for the scale factor
$a_{E}$ as function of $\widetilde{\tau}$ is actually a periodic function, so
there is more to the curve than is shown in Fig.1, but since the range for the
comoving time is already infinite in one quarter period in the space of
$\widetilde{\tau},$ the physics of the model in comoving time $0\leq t<\infty$
is already represented by the portion of the curve shown in Fig.1$.$}. This is
because the integral $\left(  \ref{physical time}\right)  $ for $t$ diverges
logarithmically when $\widetilde{\tau}$ approaches $\widetilde{\tau}_{\infty
}.$

Inflation is defined as the region $\frac{d^{2}}{dt^{2}}\left(  a_{E}\right)
>0,$ or equivalently $\frac{d}{d\tau}\left(  \frac{\dot{a}_{E}}{a_{E}}\right)
>0.$ Using $\left(  \ref{AE}\right)  $ with $\zeta=$ $\frac{1}{2},$ the
inflation region is $\widetilde{\tau}>\widetilde{\tau}_{I}\approx2.87.$ The
conversion from the conformal time to the physical time $t\left(  \tilde{\tau
}\right)  $ depends on the magnitude of the dimensionless variable $b.$ This
value of $b$ can be determined by requiring the scalar perturbation amplitude
to be of order $10^{-5}$ \cite{amplitude of P_R}. This requires $b$ to be
around $10^{-12}.$ Using this small value of $b$ we find that the time for
inflation $\widetilde{\tau}=\tilde{\tau}_{I}$ corresponds to $t_{I}%
=1.6\times10^{-38}$ seconds or $3\times10^{5}$ Planck times after the Big
Bang. Therefore, after fitting observation for the amplitude, the time of
inflation relative to the Big Bang is \textit{predicted} in our model. At the
time $t_{I}$ the energy density of the universe is $7.9\times10^{15}GeV$ which
is three orders of magnitude smaller than the Planck scale. This guarantees
the theory can\ be applied at the time of the beginning of inflation.

The analytic expression for the Hubble parameter, $H=\frac{\dot{a}_{E}}%
{a_{E}^{2}},$ that indicates the expansion rate of the universe is%

\begin{equation}
H=\left(  \frac{\kappa^{2}}{6}\right)  ^{-\frac{1}{2}}\left(  \phi_{flat}%
^{2}-s_{flat}^{2}\right)  ^{-3/2}\left(  \phi_{flat}\dot{\phi}_{flat}%
-s_{flat}\dot{s}_{flat}\right)
\end{equation}
which is plotted in Fig.2.%

\begin{center}
\includegraphics[
height=2.9291in,
width=2.9291in
]%
{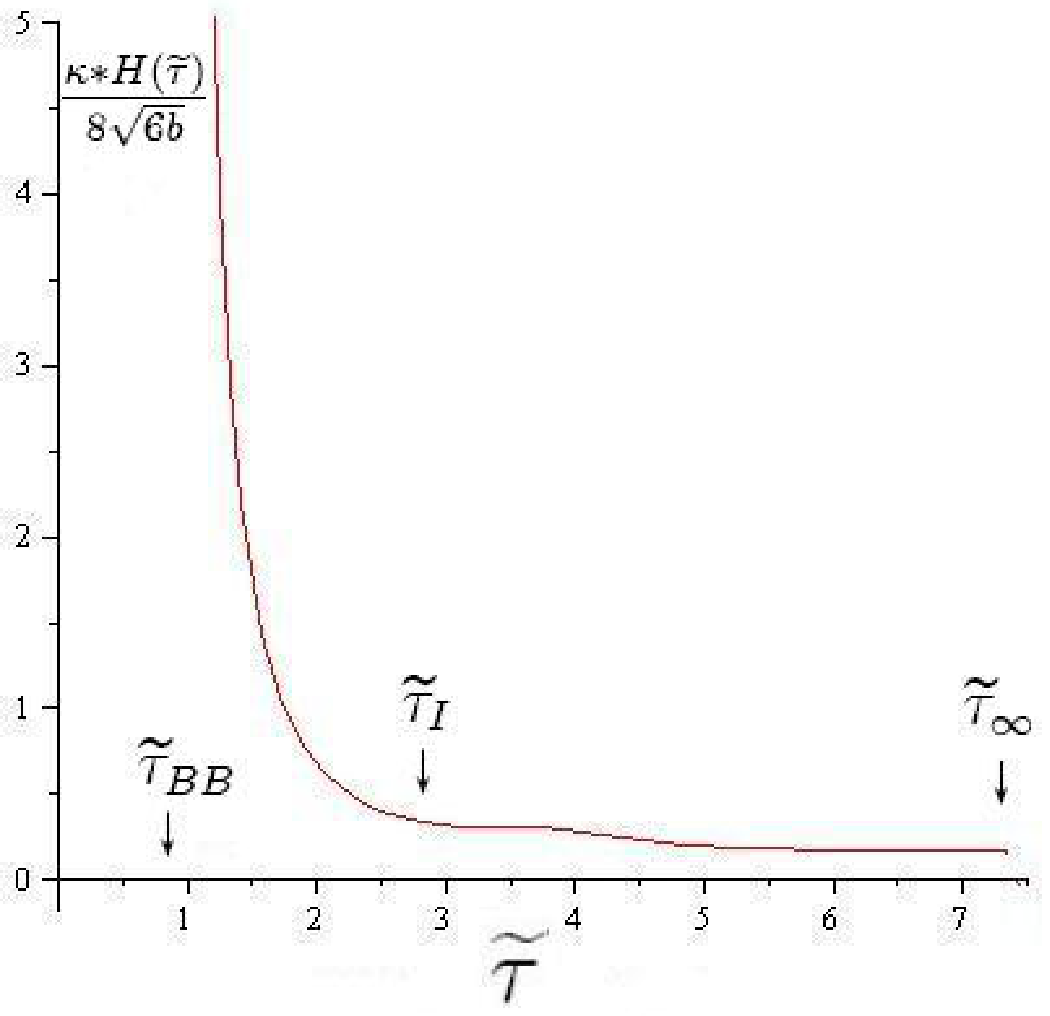}%
\\
Fig.2 - The Hubble parameter $H\left(  \tilde{\tau}\right)  $.
\label{2}%
\end{center}
From Fig.2 we can see that the Hubble parameter decreases from infinity at the
Big Bang monotonically to an asymptotic value. The Hubble rate does not change
much after inflation starts at $\widetilde{\tau}_{I}\approx2.87$, thus during
inflation we can approximate the conformal time as $\tau\approx-\frac{1}%
{a_{E}H}$, but not before.

The fact that $\frac{d}{d\tau}\left(  \frac{\dot{a}_{E}}{a_{E}}\right)  >0$
when $\widetilde{\tau}>\widetilde{\tau}_{I}$ indicates that our theory does
not have a mechanism for stopping inflation. However, like most of the
inflation theories, one follows the dynamics of the inflaton to a certain
point and claims that the inflaton decays into the Standard Model particles.
This is called reheating \cite{reheating 1}\cite{reheating2}. Similarly, in
our theory, we will assume after the theory produces enough inflation, another
aspect of the complete theory will take over. The reheating mechanism is not
the focus of this paper but certainly it is an important problem to
investigate in the future.

From Eq.(\ref{sigma as function of z and t}), we see that when the scale
factor vanishes $a_{E}\left(  \tau_{BB}\right)  =0$ the inflaton must blow up
logaritmically. Hence we find that the inflaton field $\sigma$ drops from
infinity at the Big Bang to zero in a finite conformal time and then keeps
oscillating around the potential minimum eventually dropping to $0$ at
$\widetilde{\tau}=\widetilde{\tau}_{\infty}$. Unlike most inflation theories,
this eliminates the need for declaring some arbitrary initial value for the
inflaton. As shown in Fig.3, this kind of inflation theory was never reported
before. Furthermore, most inflation theories based on slow-roll simply assume
that inflation ends before the inflaton reaches its potential minimum, but
this assumption is not valid as seen in the plot of our analytic solution.
Hence, within the slow-roll approximation, neither the connection to the Big
Bang nor the oscillation around the potential minimum could be discovered.%

\begin{center}
\includegraphics[
height=2.9283in,
width=2.9153in
]%
{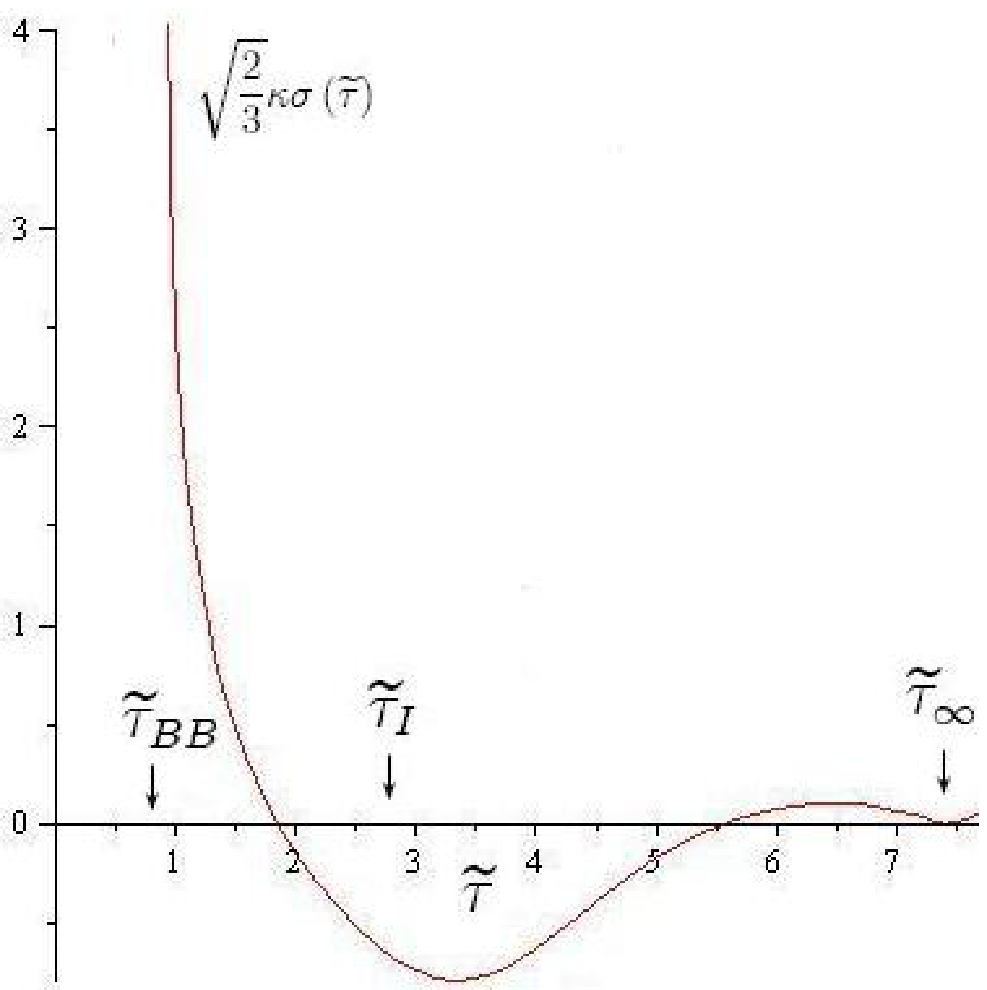}%
\\
Fig.3 - The inflaton field drops from infinity at the Big Bang and then
oscillates around the potential minimum, becoming zero at $\widetilde{\tau
}=\widetilde{\tau}_{\infty}.$%
\label{3}%
\end{center}

To calculate the power spectrum $P_{\mathcal{R}},$ the spectral index
$n_{s},n^{\prime}$ and the ratio of tensor to scalar perturbation $r$, we have
to solve the Mukhanov-Sasaki equation (\ref{muk}) with a proper boundary
condition. The solvability of this equation hinges on the properties of the
function $\ddot{z}\left(  \tau\right)  /z\left(  \tau\right)  $ whose analytic
form in our model is given by%
\begin{equation}
\frac{\overset{\cdot\cdot}{z}}{z}=\frac{\partial_{\tau}^{2}\left(  \dot
{\sigma}_{E}/H\right)  }{\left(  \dot{\sigma}_{E}/H\right)  },
\end{equation}
where $\dot{\sigma}_{E}/H$ is computed analytically by using the expressions
for $\sigma_{E}\left(  \tau\right)  ,a_{E}\left(  \tau\right)  $ in terms of
the solutions $\phi_{flat},s_{flat}$ given above as
\begin{equation}
z=\frac{\dot{\sigma}}{H}=\frac{\left(  \phi_{flat}\dot{s}_{flat}-s_{flat}%
\dot{\phi}_{flat}\right)  }{\left(  \phi_{flat}\dot{\phi}_{flat}-s_{flat}%
\dot{s}_{flat}\right)  }\left(  \phi_{flat}^{2}-s_{flat}^{2}\right)  ^{1/2}.
\label{z}%
\end{equation}
The explicit behavior of $\frac{\overset{\cdot\cdot}{z}}{z}$ in our model is
plotted in Fig.4. It seen that this is very different than the usual
approximation $\alpha/\tau^{2},$ therefore as explained earlier in the paper
following Eq.(\ref{scalar power}), we cannot use the standard approximation to
the Mukhanov-Sasaki equation.\emph{ }%
\begin{figure}
[ptb]
\begin{center}
\includegraphics[
height=3.218in,
width=3.218in
]%
{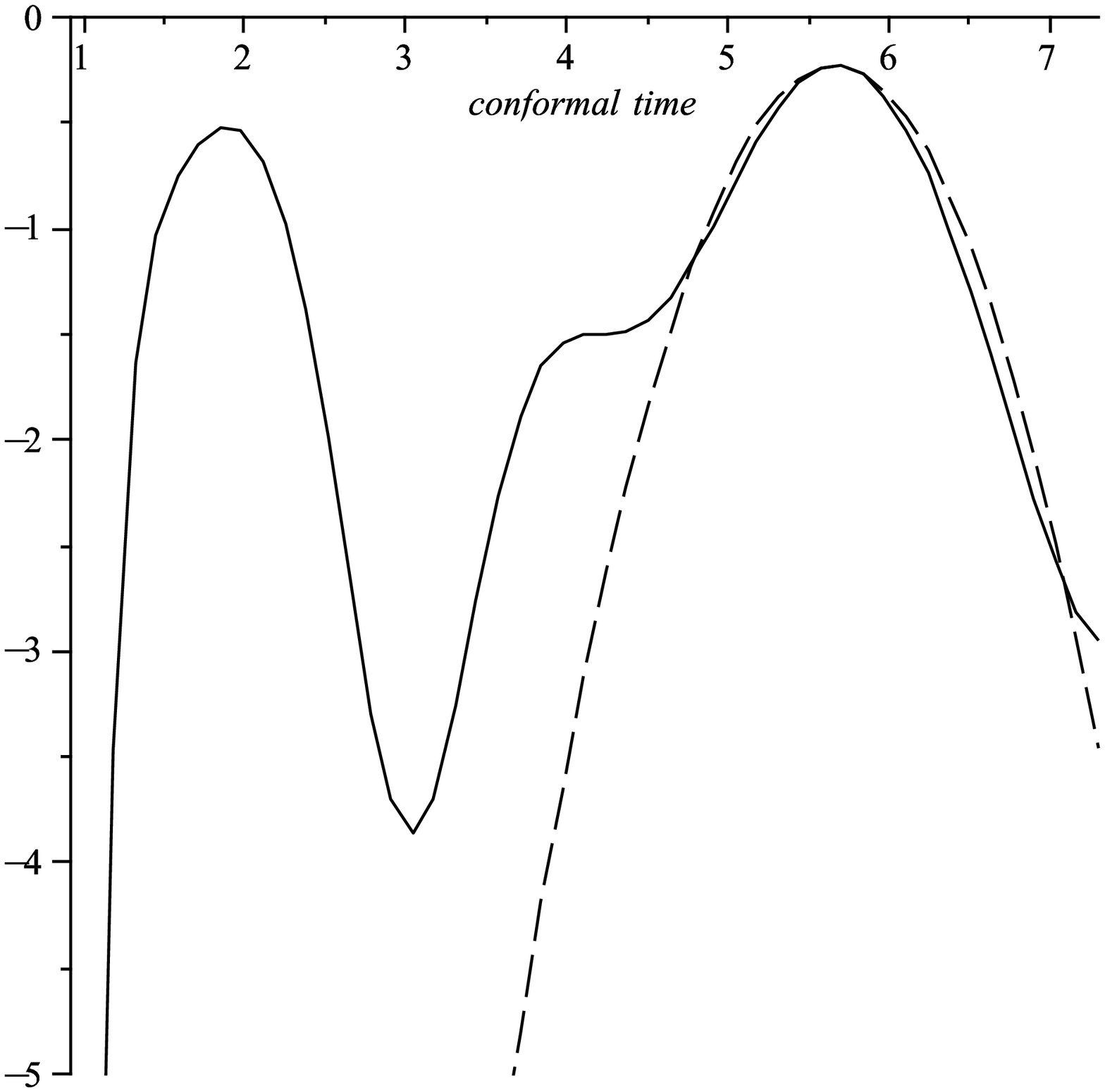}%
\caption{Fig.4 - Exact curve for $\frac{\overset{\cdot\cdot}{z}}{z}\left(
\tilde{\tau}\right)  ,$ solid; quadratic approximation, dash.}%
\end{center}
\end{figure}

To solve the Mukhanov-Sasaki equation in the vicinity of the horizon crossing
$\tau\sim\tau_{\ast}\left(  k\right)  ,$ as defined in Eq.(\ref{scalar power}%
), it is appropriate to approximate the curve $\frac{\overset{\cdot\cdot}{z}%
}{z}$ with the dotted curve shown in Fig.4. This corresponds to the upside
down dotted parabola in the figure
\begin{equation}
\frac{\overset{\cdot\cdot}{z}}{z}\approx-k_{0}^{2}-\alpha^{2}\left(  \tau
-\tau_{\max}\right)  ^{2}.
\end{equation}
We emphasize that the horizon crossing time $\tau_{\ast}\left(  k\right)  $ is
down the tail past the peak $\tau_{\ast}\left(  k\right)  >\tau_{\max}$. Here
the peak of the parabola is at time $\tau=\tau_{\max},$ while $-k_{0}^{2}$ is
the value of $\frac{\overset{\cdot\cdot}{z}}{z}$ at that time $\frac
{\overset{\cdot\cdot}{z}}{z}\left(  \tau_{\max}\right)  =-k_{0}^{2}.$ Both of
these quantities $\left(  \tau_{\max,}-k_{0}^{2}\right)  $ as well as the
parameter $\alpha^{2}$ are determined by our analytic solution for
$\frac{\overset{\cdot\cdot}{z}}{z}$ from Eq.(\ref{z}). Our analytic solution
in Eq.(\ref{flat}) depends on three parameters, namely $\left(  \frac{E}%
{b},\xi,\delta\right)  ,$ but $\frac{\overset{\cdot\cdot}{z}}{z}$ depends only
on $\left(  \xi,\delta\right)  ,$ which in turn fully determine the three
parameters of the fitting parabola. The Mukhanov-Sasaki equation then takes
the form
\begin{equation}
\overset{\cdot\cdot}{v_{k}}+\left[  \left(  k^{2}+k_{0}^{2}\right)
+\alpha^{2}\left(  \tau-\tau_{\max}\right)  ^{2}\right]  v_{k}=0.
\end{equation}
With this fitting function one can obtain analytic solutions for $v_{k}\left(
\tau\right)  $ in the desired neighborhood of $\tau<\tau_{\ast}\left(
k\right)  $ in terms of hypergeometric functions, and choose an appropriate
boundary condition at the peak of the inverted parabola at $\tau_{\max}.$

The curves shown in the various figures in this paper assume the values of
$\delta=0$ and $\xi=1/2$ as an illustrative example. For this choice of
parameters we find the fit function is given by $\frac{\overset{\cdot\cdot}%
{z}}{z}\approx-0.23-1.2\left(  \tau-5.66\right)  ^{2}.$

The details of the computation for $v_{k}\left(  \tau\right)  ,$ and the
corresponding $P_{\mathcal{R}},n_{s},n^{\prime},r$ outlined in
Eqs.(\ref{scalar power},\ref{spo2},\ref{r}), can be found in \cite{Vacuum
selection}. Below we plot the predicted spectral index $n_{s}\left(
\tilde{\tau}\right)  $, running of the spectral index $n^{\prime}\left(
\tilde{\tau}\right)  $, as well as constrain the tensor to scalar ratio
$r\left(  \tilde{\tau}\right)  ,$ as follows (where we recall $\tilde{\tau
}=2\left\vert cE\right\vert ^{1/4}\tau$)%

\begin{center}
\includegraphics[
height=1.7815in,
width=3.3797in
]%
{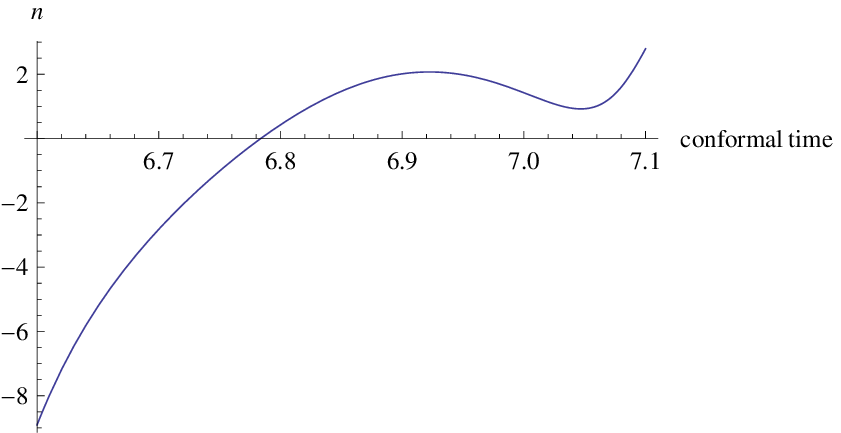}%
\\
Fig.5 - The spectral index $n_{s}\left(  \tilde{\tau}\right)  $. At
$\tilde{\tau}=7.05$ the value is 0.97.
\end{center}

The observational data restricts the value of the spectral index $n_{s}$ to
between $0.92$ to $1$ at 60 e-folds before the end of inflation, and the
tensor to scalar ratio between $0$ and $0.5.$ These ranges are the two-sigma
region which corresponds to $95\%$ confidence limit from the WMAP 7 years data
\cite{Kinney WMAP 5years}. Since in our theory, inflation has to be truncated
by hand, we just need to know if there is a time during inflation such that
the values of $n_{s}$ and $r$ reside within this range. If such a time exists,
we can let inflation develop for 60 more e-folds and truncate the theory.

From the figures we see that at $\widetilde{\tau}\approx7.05,$ or equivalently
$t=3\times10^{-37}$ seconds after the Big Bang, the value of $n_{s}$ is around
0.97$.$%

\begin{center}
\includegraphics[
height=1.8092in,
width=3.3797in
]%
{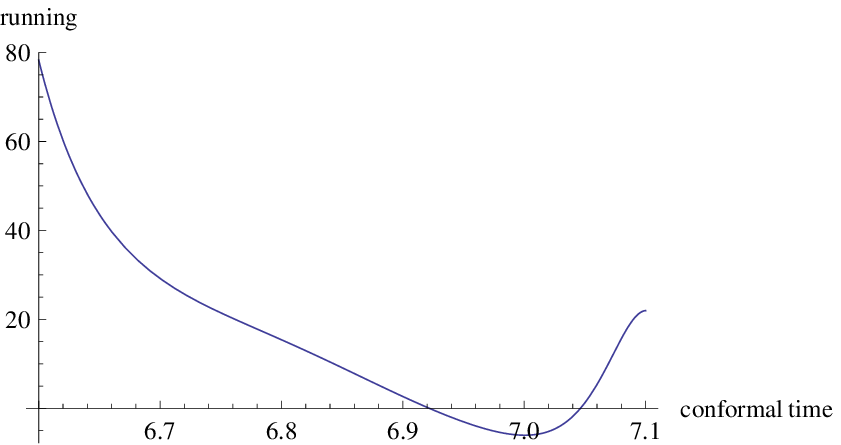}%
\\
Fig.6 - The running of the spectral index $n^{\prime}\left(  \tilde{\tau
}\right)  $. At $\tilde{\tau}=7.05$ its value is close to zero.
\end{center}

The analytic expression for $n^{\prime}$ is not very transparent. It is
plotted in Fig.6$.$ We can see that at $\widetilde{\tau}\approx7.05$ the
running is around $0.$ This is within the upper bound of the WMAP 7 years data
estimates of around $0.05$.

Meanwhile, the observational upper bound of the tensor to scalar ratio $r$ is
satisfied in our model and merely constrains the boundary condition of tensor
mode perturbation equation at $\tau_{\max}$ \cite{Vacuum selection}\emph{.}

\section{Conclusion and future directions}

\label{conclusion}

In this paper we have established a definite relation between the Big Bang and
inflation, where neither one could be considered as separate events due to
separate physical reasons. This is a new point of view for the cosmological
evolution of our universe.

We have also presented both a new method for solving exactly certain
scalar-tensor theories, as well as a sufficiently realistic new model of
inflation. The theory has very few parameters, but yet it makes
phenomenologically consistent interesting predictions based on an analytic
solution, and simultaneously reveals several new physical features not
discussed before.

The method consists of enlarging the action to a form with local Weyl
symmetry, with some special potential. Both of these aspects were inspired by
2T-physics. We can then solve the theory in a gauge where exact solutions are obtained.

This method should be regarded as one of the useful technical by products of
2T-gravity, while the physics content is also one of the novel predictions of
2T-physics. The new features of our solution could have been obtained within
1T-physics, but was not, thus demonstrating that 1T-physics lacks the guidance
that 2T-physics supplies systematically in the form of hidden dualities and
hidden symmetries as explained in \cite{GMtalk}\cite{Dual field theories}%
\cite{Dual fild theories fermion}.

We could in principle use the same techniques to obtain exact solutions for
more general inflation models with more than one scalar field, and more
complicated potentials. As we have seen in this paper, a theory that appears
intractable may be presented as a complicated 1T-physics shadow of a
2T-physics theory, while in another shadow that corresponds to a more
tractable gauge choice, the theory is handled much more easily and even
solvable. As illustrated in \cite{GMtalk}\cite{Dual field theories}\cite{Dual
fild theories fermion} 2T-physics offers this kind of new insights that are
not available in 1T-physics. In this paper we have essentially used this
duality idea of 2T-physics for a specific case in the context of 2T-gravity,
but we expect more general applications of this concept in future work.

The exact solutions obtained with our methods allowed us to analyze the theory
more precisely than using the slow-roll approximation. This revealed
properties of the theory that could not be found by slow-roll analyses, such
as the predictions for the time delay between the Big Bang and inflation or
the oscillatory behavior of the inflaton.

Satisfying the constraints coming from the first order phenomenological
parameters, the spectral index, the running of the specrtral index and the
amplitude of scalar power spectrum, we have constructed a phenomenological
model, with the action (\ref{theory}) and inflaton potential%
\begin{equation}
V\left(  \sigma\right)  =\left(  \frac{6}{\kappa^{2}}\right)  ^{2}b\left(
64\sinh^{4}\left(  \sqrt{\frac{\kappa^{2}}{6}}\sigma\right)  +\cosh^{4}\left(
\sqrt{\frac{\kappa^{2}}{6}}\sigma\right)  \right)  ,
\end{equation}
where $b$ is a positive dimensionless parameter of order $10^{-12}.$ A small
parameter like $b$ is common in general inflation theories. This makes the
coefficient of the potential $V$ of order $\left(  \frac{6}{\kappa^{2}%
}\right)  ^{2}b\sim\left(  6\times10^{15}\text{GeV}\right)  ^{4},$ revealing
perhaps an interesting scale since it is not too far from the grand
unification (GUT) scale.

This particular theory has several interesting behaviors. However, it does not
predict when inflation ends, which is one of its weak points. In a complete
theory, a modification around the GUT scale mentioned above could provide the
desired mechanism to end inflation.

It is also interesting to note that we obtain a cyclic cosmology, not unlike
ref.\cite{cyclicST}, under the following conditions
\begin{equation}%
\begin{array}
[c]{l}%
\text{1. A special value of the integration constant }\delta\text{ given by
}\\
\text{ \ \ the quarter period of the elliptic function }cn\left(  \tilde{\tau
}+\delta\right)  .\\
\text{2. A quantized value of the parameter }\zeta=\frac{2}{n},\text{ with
integer }n.
\end{array}
\end{equation}
Under these conditions our solution describes an universe that expands from a
big bang and shrinks to a big crunch, and repeats this cycle indefinitely. In
the meantime the equation of state $w$
\begin{equation}
w\equiv\frac{\dot{\sigma}_{E}^{2}/2a_{E}^{2}-V\left(  \sigma_{E}\right)
}{\dot{\sigma}_{E}^{2}/2a_{E}^{2}+V\left(  \sigma_{E}\right)  }%
\end{equation}
never goes below $-1$. In this case the effective gravitational
\textquotedblleft constant\textquotedblright\ $\left(  \phi_{flat}^{2}\left(
\tau\right)  -s_{flat}^{2}\left(  \tau\right)  \right)  $ that appears in the
action Eq.(\ref{action}) never changes sign in any cycle although it becomes
zero at each big bang or big crunch. This cyclic cosmological scenario will be
discussed in more detail separately in a future paper \cite{cyclic universe}.

The applications of our equations\ are not restricted to inflation theories.
They might also be useful to discuss dark energy to explain the current
expansion of the universe. However, this will require a fine tuning of our
parameters$.$ If one wants to use a similar model for dark energy, $b$ has to
be as small as $10^{-120}.$ This is because the current Hubble time is way too
big compared to the Planck time. Including supersymmetry in our approach may
play a role to suppress the value of emergent cosmological constant at late
times (see \cite{2tsugra}). A modified version of our theory might be
extendible up to the current era, and then be applicable also to the physics
of dark energy.

\begin{acknowledgments}
We thank Elena Pierpaoli, Loris Colombo, Nicholas Warner, Yueh-Cheng Kuo, and
Guillaume Qu\'{e}lin at USC for helpful discussions
\end{acknowledgments}

\appendix{}

\section{Properties of Jacobi elliptic
functions\label{Jacobi elliptic functions}}

The Jacobi elliptic functions used in this paper are $sn\left(  z|m\right)
,cn\left(  z|m\right)  $ and $dn\left(  z|m\right)  .$ There are two
parameters $z$ and $m$ for each of them. The parameter $m$ determines the
period under translations of the parameter $z.$ The period is given by
\begin{equation}
period=4\ast\int_{0}^{\frac{\pi}{2}}\frac{d\theta}{\left(  1-m\sin^{2}%
\theta\right)  ^{\frac{1}{2}}}.
\end{equation}
With $m=0$ the period for all the Jacobi elliptic functions is $2\pi.$ For
$m=0$ these functions reduce to the familiar sine and cosine, $sn\left(
z|0\right)  =\sin\left(  z\right)  $, $cn\left(  z|0\right)  =\cos\left(
z\right)  $, and then $dn\left(  z\right)  =1.$ In our solutions we have
$m=\frac{1}{2}$. This results in the period to be about $7.42$ in $z,$ but
note that $z$ is related by a factor to $\widetilde{\tau}$ or it is translated
by $\delta$ in the various entries in Tables 1,2,3. The functions $sn\left(
z|\frac{1}{2}\right)  ,cn\left(  z|\frac{1}{2}\right)  $ and $dn\left(
z|\frac{1}{2}\right)  $ are plotted in Fig.7
\begin{center}
\includegraphics[
height=2.9058in,
width=2.9058in
]%
{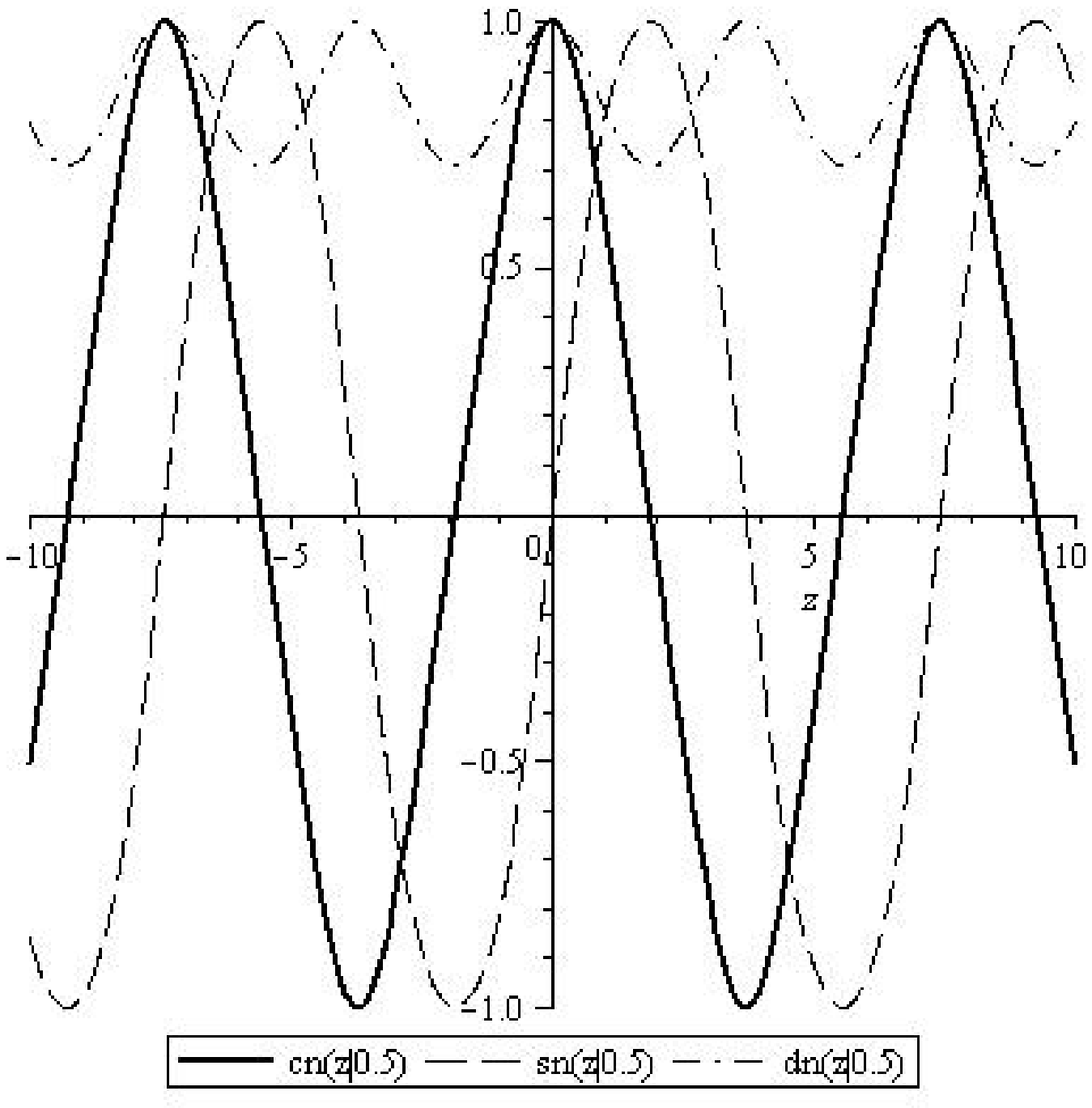}%
\\
Fig.7 Plots of cn$\left(  z|\frac{1}{2}\right)  $ (solid), sn$\left(
z|\frac{1}{2}\right)  $ (dash) and dn$\left(  z|\frac{1}{2}\right)  $
(dash-dot).
\label{7}%
\end{center}
We can see that the behavior of $sn$ and $cn$ are very similar to their
counterpart trigonometric functions. They also satisfy properties similar to
trigonometric functions, such as%
\begin{equation}
\left(  sn\left(  z|m\right)  \right)  ^{2}+\left(  cn\left(  z|m\right)
\right)  ^{2}=1;\;\ m\left(  sn\left(  z|m\right)  \right)  ^{2}+\left(
dn\left(  z|m\right)  \right)  ^{2}=1
\end{equation}

The derivative of Jacobi elliptic functions are given in terms of expressions
somewhat similar to those for trigonometric functions.%
\begin{align*}
\frac{d}{dz}sn\left(  z|m\right)   &  =cn\left(  z|m\right)  \times dn\left(
z|m\right)  ,\\
\frac{d}{dz}cn\left(  z|m\right)   &  =-sn\left(  z|m\right)  \times dn\left(
z|m\right)  ,\\
\frac{d}{dz}dn\left(  z|m\right)   &  =-m\times sn\left(  z|m\right)  \times
cn\left(  z|m\right)  .
\end{align*}
One can look up more properties of Jacobi elliptic functions at \cite{Hand
book}.

\end{document}